\documentclass[graybox]{svmult}

\usepackage{type1cm}        
\usepackage{makeidx}         
\usepackage{graphicx}        
\usepackage{multicol}        
\usepackage[bottom]{footmisc}

\usepackage{newtxtext}       %
\usepackage[varvw]{newtxmath}       

\makeindex             

\usepackage[authoryear]{natbib}
\usepackage{aas_macros}
\usepackage{hyperref}

\begin{document}

\title*{Climates of Terrestrial Exoplanets and Biosignatures}
\author{
Siddharth Bhatnagar*$^\star$\orcidID{0000-0001-9178-168X},
Emeline Bolmont*$^\star$\orcidID{0000-0001-5657-4503},
Nikita J. Boeren*\orcidID{0000-0001-6162-6953},
Janina Hansen*\orcidID{0009-0003-1247-8378},
Björn Konrad*\orcidID{0000-0002-9912-8340},
Leander Schlarmann*\orcidID{0000-0001-5800-132X},
Eleonora Alei\orcidID{0000-0002-0006-1175},
Marie Azevedo\orcidID{0009-0009-5904-751X},
Marrick Braam\orcidID{0000-0002-9076-2361},
Guillaume Chaverot\orcidID{0000-0003-4711-3099},
Jonathan Grone\orcidID{0000-0001-5074-265X},
Kaustubh Hakim\orcidID{0000-0003-4815-2874},
Mathilde Houelle\orcidID{0009-0001-7217-4099},
Daniel Kitzmann\orcidID{0000-0003-4269-3311},
Christophe Lovis\orcidID{0000-0001-7120-5837},
Antoine Pommerol\orcidID{0000-0002-9165-9243},
Sascha P. Quanz\orcidID{0000-0003-3829-7412},
Martin Turbet\orcidID{0000-0003-2260-9856},
Audrey Vorburger\orcidID{0000-0002-7400-9142},
Susanne F. Wampfler\orcidID{0000-0002-3151-7657},
Francis Zong Lang\orcidID{0009-0005-3162-3694}
}

\institute{Siddharth Bhatnagar \at Department of Astronomy, Universit\'e de Gen\`eve, Chemin Pegasi 51, CH-1290 Versoix, Switzerland, \email{siddharth.bhatnagar@unige.ch}
\at Centre pour la Vie dans l'Univers, Universit\'e de Gen\`eve, Gen\`eve, Switzerland
\and Emeline Bolmont \at Department of Astronomy, Universit\'e de Gen\`eve, Chemin Pegasi 51, CH-1290 Versoix, Switzerland, \email{emeline.bolmont@unige.ch}
\at Centre pour la Vie dans l'Univers, Universit\'e de Gen\`eve, Gen\`eve, Switzerland
\and Nikita J. Boeren \at Space Research and Planetary Sciences, Physics Institute, University of Bern, 3012 Bern, Switzerland, \email{nikita.boeren@unibe.ch}
\and Janina Hansen \at ETH Zurich, Institute for Particle Physics \& Astrophysics, Wolfgang-Pauli-Str. 27, 8093 Zurich, Switzerland, \email{jahansen@phys.ethz.ch}
\and Björn Konrad \at ETH Zurich, Institute for Particle Physics \& Astrophysics, Wolfgang-Pauli-Str. 27, 8093 Zurich, Switzerland, \email{konradb@student.ethz.ch}
\and Leander Schlarmann \at Space Research and Planetary Sciences, Physics Institute, University of Bern, \email{leander.schlarmann@unibe.ch}
\and Eleonora Alei \at NPP Fellow, NASA Goddard Space Flight Center, 8800 Goddard Rd, Greenbelt, 20771, MD, USA, \email{eleonora.alei@nasa.gov}
\and Marie Azevedo \at Space Research and Planetary Sciences, Physics Institute, University of Bern, \email{marie.azevedo@unibe.ch}
\at Group of Applied Physics and Institute for Environmental Sciences, Universit\'e de Gen\`eve, Gen\`eve, Switzerland
\and Marrick Braam \at Center for Space and Habitability, University of Bern \email{marrick.braam@unibe.ch}
\and Guillaume Chaverot \at Univ. Grenoble Alpes, CNRS, IPAG, 38000 Grenoble, France, \email{guillaume.chaverot@univ-genoble-alpes.fr}
\and Jonathan Grone \at Center for Space and Habitability, University of Bern \email{jonathan.grone@unibe.ch}
\and Kaustubh Hakim \at KU Leuven, Institute of Astronomy, Celestijnenlaan 200D, 3001 Leuven, Belgium, \email{kaustubh.hakim@kuleuven.be}
\at Royal Observatory of Belgium, Ringlaan 3, 1180 Brussels, Belgium
\and Mathilde Houelle \at Department of Astronomy, Universit\'e de Gen\`eve, Chemin Pegasi 51, CH-1290 Versoix, Switzerland, \email{mathilde.houelle@unige.ch}
\at Centre pour la Vie dans l'Univers, Universit\'e de Gen\`eve, Gen\`eve, Switzerland
\and Daniel Kitzmann \at Space Research and Planetary Sciences, Physics Institute, University of Bern, \email{daniel.kitzmann@unibe.ch}
\and Christophe Lovis \at Department of Astronomy, Universit\'e de Gen\`eve, Chemin Pegasi 51, CH-1290 Versoix, Switzerland, \email{christophe.lovis@unige.ch}
\and Antoine Pommerol \at Space Research and Planetary Sciences, Physics Institute, University of Bern, \email{antoine.pommerol@unibe.ch}
\and Sascha P. Quanz \at ETH Zurich, Institute for Particle Physics \& Astrophysics, Wolfgang-Pauli-Strasse 27, 8093 Zurich, Switzerland, \email{sascha.quanz@phys.ethz.ch}
\at ETH Zurich, Department of Earth and Planetary Sciences, Sonneggstrasse 5, 8092 Zurich, Switzerland
\and Martin Turbet \at Laboratoire de M\'et\'eorologie Dynamique/IPSL, CNRS, Sorbonne Universit\'e, Ecole Normale Sup\'erieure, Universit\'e PSL, Ecole Polytechnique, Institut Polytechnique de Paris, 75005 Paris, France, \email{martin.turbet@lmd.ipsl.fr}
\at Laboratoire d'astrophysique de Bordeaux, Univ. Bordeaux, CNRS, B18N, all\'ee Geoffroy Saint-Hilaire, 33615 Pessac, France
\and Audrey Vorburger \at Space Research and Planetary Sciences, Physics Institute, University of Bern, \email{audrey.vorburger@unibe.ch}
\and Susanne F. Wampfler \at Center for Space and Habitability, University of Bern, Gesellschaftsstrasse 6, CH-3012 Bern, Switzerland, \email{susanne.wampfler@unibe.ch}
\and Francis Zong Lang \at Center for Space and Habitability, Physics Institute, University of Bern, \email{francis.zong@unibe.ch}
}

%

\maketitle

\section{Introduction}


Thirty years after the detection of the hot Jupiter 51 Pegasi b \citep{1995Natur.378..355M}, there are now more than 5800 exoplanets discovered (\url{https://exoplanetarchive.ipac.caltech.edu/}). 
Among those 5800 planets, and due to instrumental biases, a majority of them are hotter and bigger than our Earth.
However, thanks to instrumental developments, we are now pushing the detection limit towards smaller and colder planets.
Although several instruments (will) aim to search for an Earth-like planet around a Sun-like star, for instance with the ESPRESSO spectrograph \citep{2021A&A...645A..96P} and later with PLATO \citep{2014ExA....38..249R}, an Earth 2.0 remains elusive. 
Once such planets are detected, they will be targets of telescopes like the Habitable Worlds Observatory (HWO) and the Large Interferometer For Exoplanets (LIFE, see Section~\ref{sec:biosignatures}). 
In the meantime, we now detect small temperate planets mostly around M-dwarfs, such as Proxima-b \citep{2016Natur.536..437A} or the TRAPPIST-1 planets \citep{2017Natur.542..456G}. 
However, these planets are very different from our Earth in so far as (but not limited to): 1) their host star itself is very different from our Sun with a spectral energy distribution shifted to higher wavelengths and is generally more active than the Sun; 2) their rotation is thought to be tidally locked offering one permanent day side and one permanent night side.
One aspect which makes these planets particularly interesting is that their atmosphere can be already probed by transmission/emission spectroscopy or emission by the JWST \citep[e.g.][for the TRAPPIST-1 planets]{2023Natur.618...39G,2023Natur.620..746Z,2023ApJ...955L..22L,2025arXiv250902128G} and will also later be the target of future instruments such as RISTRETTO at the VLT and ANDES/PCS at the ELT (see Section~\ref{sec:biosignatures}).

Having detected small temperate planets, the question of their habitability and whether we could ever find {signs of }life on them is a powerful driver in the community.
Habitability refers to the potential of a global (or even local) environment to support life (see Section~\ref{sec:habitability}). 
{Summarised from \citet{2018haex.bookE..57M} and \citet{turbet2018thesis}, habitability can be} constrained by five key environmental factors: (i) the presence of liquid water, (ii) a source of energy, (iii) essential nutrients and building blocks (CHNOPS: carbon, hydrogen, nitrogen, oxygen, phosphorus, and sulfur), (iv) protection from harmful conditions, and (v) long-term stability. 
Since exoplanets are difficult to assess due to their vast distances, the planets and moons within our Solar System serve as crucial proxies for studying climates, habitability, and biosignatures. 
A key aspect of the search for biosignatures-both within our Solar System and beyond-is understanding the fundamental requirements for life, allowing us to determine where to look. 
Many of the factors determining habitability can be assessed either in situ or remotely, such as by analyzing a planet's chemical composition (including its atmosphere) and internal structure to detect the presence of liquid water. 
Within our Solar System, several targets have been identified as highly promising for habitability. 
The most notable among them are Mars, Jupiter's moon Europa, and Saturn's moon Enceladus, which rank highest in terms of their potential to support life.

\begin{figure}[ht!]
     \centering
     \includegraphics[width=\linewidth]{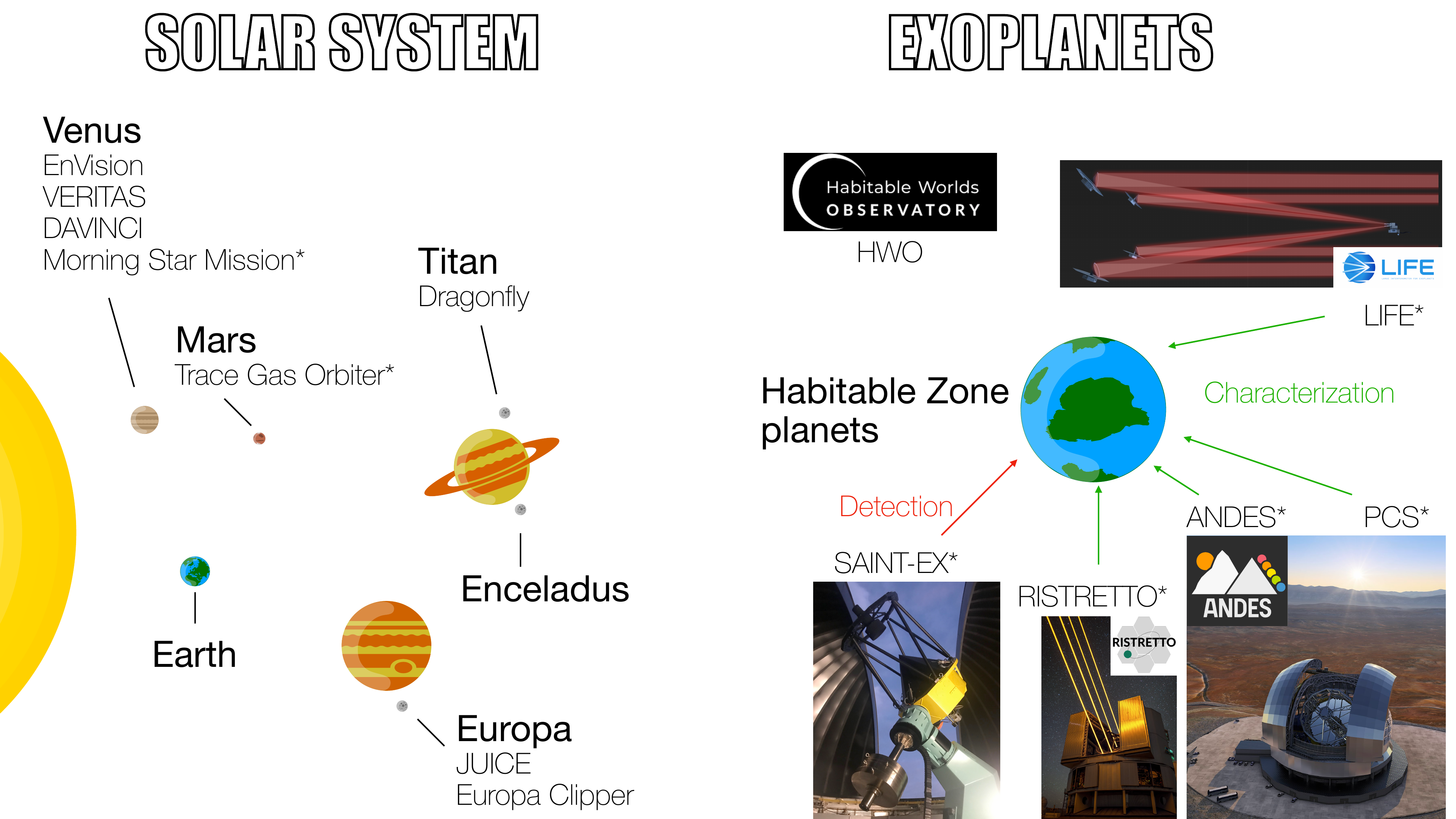}
     \caption{The different Solar System bodies and exoplanets this chapter focuses on, and associated missions. Those with a strong PlanetS involvement are shown with an asterisk.}
     \label{fig:SS_Exo}
 \end{figure}

This chapter aims at providing a non-exhaustive overview of this science topic with a focus on the science done within the {NCCR} PlanetS during the period 2014-2025.
Section~\ref{sec:habitability} reviews the concept of habitability both for solar system objects and exoplanets and how it is accessed.
Section~\ref{sec:biosignatures} offers a discussion of the concept of biosignatures and how we can proceed to detect them. Figure~\ref{fig:SS_Exo} illustrates the topics and objects mentioned in this chapter.

\newcommand{\SubItem}[1]{
    {\setlength\itemindent{15pt} \item[--] #1}
}

\newcommand{\SubSubItem}[1]{
    {\setlength\itemindent{30pt} \item[*] #1}
}

\section{Climate Simulations and Habitability}\label{sec:habitability}

Behind the concept of habitability is the idea of finding {signs of }extraterrestrial life, or rather environments conducive of life. 
However, as of today, we only know about one planet in the whole universe which hosts life: our own planet Earth. 
The fact that we have only one data point leads us to pragmatically consider a life ``as we know it'' in our search for {signs of }extraterrestrial life. 
This motivates the definition of ``habitability'' given earlier.
In the requirements for habitability, liquid water is considered the limiting factor based on our experience in the solar system.
Liquid water has many interesting properties: It is liquid for a wide range of pressures and temperatures (and relatively warm temperatures), it acts as an efficient solvent for organic chemistry, it also allows for hydrogen bonds which help stabilize big organic macromolecules.
Finally, it is also thought that a surface is necessary as it ensures the building blocks (CHNOPS) and minerals can be available in numbers.

\begin{figure}[ht!]
     \centering
     \includegraphics[width=\linewidth]{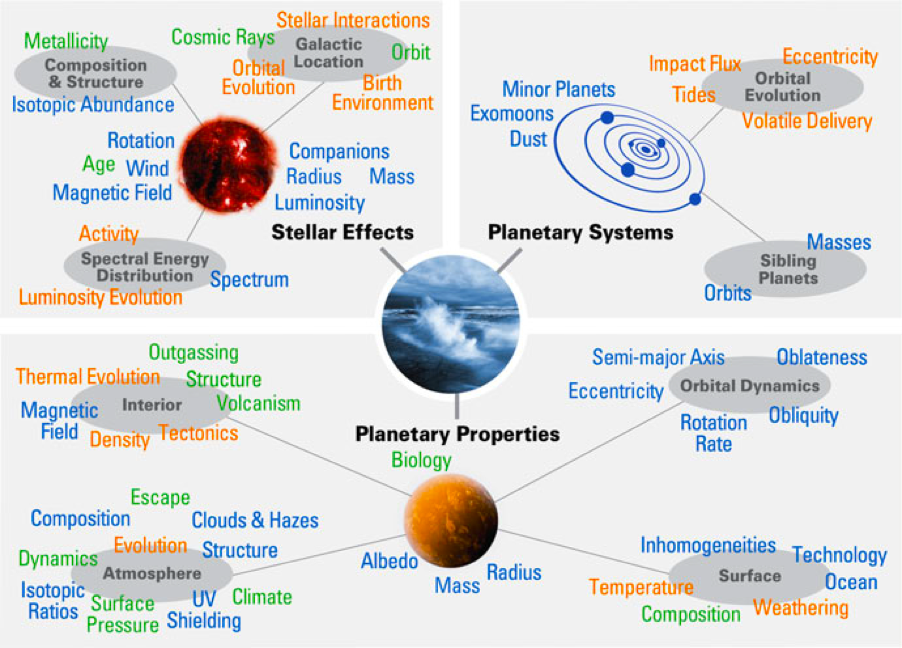}
     \caption{The various factors affecting planetary evolution and habitability.{ Font colors indicate observability: blue for directly detectable features with sufficiently powerful telescopes, green for those requiring model-based interpretation and orange for properties accessible mainly through theory.} Credits: \citep{2018haex.bookE..57M}}
     \label{fig:meadows_barnes_2018}
 \end{figure}

The presence of liquid water is only possible in a set of planet characteristics such as its mass and radius, the presence and composition of an atmosphere and its distance from the star.
Not only the planet's properties are important but also those of the host star and the planetary system the planet is part of \citep[e.g.][]{2018haex.bookE..57M} (see Figure~\ref{fig:meadows_barnes_2018} for an overview).

The concept of Habitability 
gave birth to the concept of the Habitable Zone \citep{1993Icar..101..108K,2007A&A...476.1373S}. 
It is used to estimate the habitability potential of a planet and the amount of potentially habitable planets in the galaxy.
The Habitable Zone (HZ) is the region around a star where a planet can potentially host water in a liquid phase.


The inner edge of the habitable zone is defined by the runaway greenhouse \citep{1993Icar..101..108K} or moist greenhouse limit \citep{2015ApJ...813L...3K}. 
As temperatures rise, more water vapor enters the atmosphere, amplifying warming until equilibrium is reached. 
If the stratosphere host enough water vapour \citep{1984Icar...57..335K}, it can be photolysed and escape to space, leading to water loss and potential habitability loss (moist greenhouse limit, though {we debated its existence in \citealt{2023A&A...680A.103C}}).
In the runaway greenhouse scenario, increasing water vapor traps outgoing thermal radiation, causing unchecked heating until all surface water evaporates. 
Cooling resumes only when thermal emission shifts into water vapor opacity windows in the visible.

The outer edge of the habitable zone is defined by the maximum greenhouse limit \citep{1993Icar..101..108K}, beyond which adding CO$_2$ no longer increases surface warming. Instead, rising CO$_2$ levels enhance Rayleigh scattering, increasing planetary albedo and reducing the net greenhouse effect. This framework is based on Earth's carbonate-silicate cycle \citep{Walker_1981}, which regulates CO$_2$ via silicate weathering and volcanic outgassing \citep[e.g.][]{1988Icar...74..472K}. As weathering is more efficient in warm climates, the cycle acts as a climate stabilizer, maintaining habitability over billions of years and enabling recovery from snowball episodes. Farther from the star, CO$_2$ accumulates if this cycle is active, potentially extending habitability. However, cold-trapping of CO$_2$ can hinder deglaciation \citep{2017E&PSL.476...11T}, and excessive CO$_2$ (several bars) raises albedo, offsetting warming \citep{2013ApJ...765..131K}. Around low-mass stars, Rayleigh scattering is weaker, but {we showed that} CO$_2$ collapse on the nightside becomes a dominant constraint \citep{2018A&A...612A..86T}.

While the HZ is a useful concept, it has its limitations. 
Since stellar luminosity evolves over time, the HZ boundaries shift, leading to the concept of the Continuous Habitable Zone (CHZ) \citep{1978Icar...33...23H}, where a planet remains within the HZ throughout its star’s main sequence lifetime.
The assumption that carbonate-silicate cycling stabilizes the CHZ can break down at the thermodynamic limit of weathering. 
At lower surface temperatures than typically assumed, {we showed that }silicate weathering may switch from a negative to a positive feedback, destabilizing the climate system \citep{2021PSJ.....2...49H}.
The inner edge of the HZ also depends on the rotation of the planet, with slow rotators being able to sustain habitable conditions closer to the star than fast rotators \citep{2013ApJ...771L..45Y,2014ApJ...787L...2Y,2020JGRE..12506276W}.
Additionally, the HZ limits are highly dependent on the composition of the atmosphere. 
For instance, not taking into account a possible increase of greenhouse gases due to the carbonate-silicate cycle when a planet is farther from the star would mean that the HZ outer limit is much closer in that what is computed for the ``maximum greenhouse limit''. 
For an Earth-like atmospheric composition, reducing the solar input by even 6-9\% is enough to push the planet into a snowball state \citep[e.g.][]{voigt2010transition}.
The background gas has also been shown to have an influence on the inner edge of the HZ \citep[e.g][]{2013NatGe...6..661G,2022A&A...658A..40C}. 
{Likewise, it was established that hydrogen-dominated atmospheres can extend the outer edge compared to an Earth-like case, thanks to Collision Induced Absorption}
\citep[CIA; e.g.][]{2011ApJ...734L..13P,2017ApJ...837L...4R}. 
{Building on these studies, we demonstrated that CIA can significantly impact habitable conditions \citep[][see more in Section~\ref{sub:Habitability_Exo}]{2022NatAs...6..819M,2025Life...15...79K}.}
By definition, the HZ is calculated for a given composition, but as for the Earth, the atmospheric composition can vary quite dramatically during the lifetime of a planet \citep[e.g.][]{2014Natur.506..307L}. 
The atmospheric evolution of a rocky planet is deeply linked with the interior evolution and tectonic regime \citep{2015ApJ...812...36F,2019A&A...627A..48H}. 
Planets around stars of lower mass than the Sun can also be subject to quite efficient atmospheric escape, which can also dramatically change the extent and composition of the atmosphere \citep{2020JGRA..12527639G,2022ApJ...933..115K}. 
If this process is too extreme, planets can lose their atmosphere altogether (giving rise to the ``cosmic shoreline'' concept, e.g. \citealt{2025ApJ...987...22C}).
Nonetheless, {in PlanetS, we showed that }larger rocky planets have a higher tendency to hold on to their atmosphere than their smaller counterparts \citep{2024SpScT...4...75G}.
The discovery of rocky planets around low mass stars poses another challenge for habitability.
Many of these planets are theorised to be tidally locked, with permanent day and nightsides.
The temperatures on the nightside could be low enough for the major component of the atmosphere to condense, leading to "atmospheric collapse" \citep[e.g.][]{2015ApJ...806..180W}
{In \citet{2020A&A...638A..77A}, we contributed to that question by developing new simplified analytical models that capture the key physical mechanisms behind heat redistribution and collapse pressure, thereby complementing the numerical GCM-based constraints of \citet{2015ApJ...806..180W}.}
However, tidal locking has also been shown to enable planets to retain surface liquid water at higher stellar irradiations than would be possible for rapidly rotating planets \cite{2013ApJ...771L..45Y,2020JGRE..12506276W}.

This section examines these aspects, {detailing further} PlanetS{' contributions in the research} on habitability.
Section~\ref{sub:Tools_habitability} presents the tools developed and utilized within PlanetS, while Sections~\ref{sub:Habitability_SS} and \ref{sub:Habitability_Exo} summarize key scientific findings in Solar System studies and exoplanet research, respectively.


\subsection{Tools to study habitability}\label{sub:Tools_habitability}



To explore planetary habitability, a range of numerical tools have been developed - chief among them being climate models. 
These simulate essential atmospheric properties such as temperature, pressure, and composition, and vary widely in complexity. Climate models serve two main purposes: (1) to simulate the atmospheric and/or oceanic state of a planet, enabling assessments of its habitability, thermal structure, and cloud dynamics using a hierarchy of models—including 1-D radiative-convective models (Sec.\ref{subsub:1DRCEs}), 1-D energy balance models (Sec.\ref{subsub:1DEBM}), and 3-D general circulation models (Sec.\ref{subsub:3D_models}); and (2) to perform atmospheric retrievals, using observed spectra or thermal emission to infer plausible atmospheric structures and compositions (Sec.\ref{subsub:retrievals}). 

\begin{figure}[ht!]
     \centering
     \includegraphics[width=\linewidth]{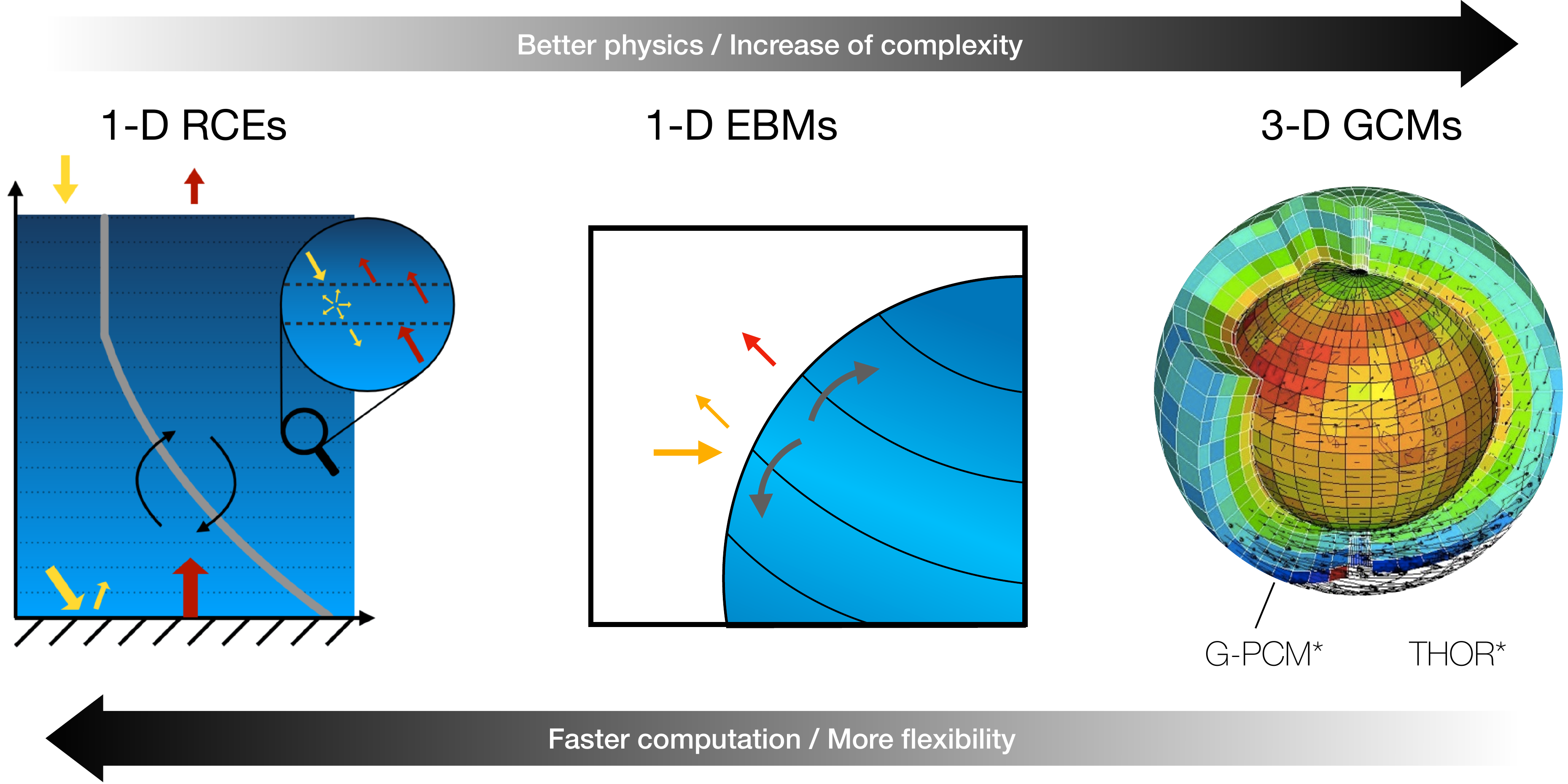}
     \caption{The various climate models used in the PlanetS community.}
     \label{fig:Fig_tools}
 \end{figure}

{Increasing complexity inevitably trades off against computation speed and the number of required input parameters and assumptions. Simple models (\ref{subsub:1DRCEs} and \ref{subsub:1DEBM}) allow for rapid exploration of wide parameter spaces with relatively few inputs but rely on parametrising a larger fraction of the underlying physics. In contrast, more comprehensive models (\ref{subsub:3D_models}) capture many physical processes self-consistently but introduce numerous detailed parameterisations and additional degrees of freedom that can complicate interpretation. Balancing complexity and parameterisation is therefore a central consideration when selecting the appropriate tool.} PlanetS has played a central role in advancing both the development and application of this full hierarchy of climate models (see Figure~\ref{fig:Fig_tools}).

\subsubsection{1-D radiative-convective models}\label{subsub:1DRCEs}

Radiative-convective models (RCEs) were the first climate models developed to study Earth’s atmosphere \citep[e.g.][]{1967JAtS...24..241M} and remain widely used today due to their simplicity, flexibility, and computational efficiency. 
They have been instrumental in estimating the boundaries of the Habitable Zone \citep[e.g.][]{1993Icar..101..108K,2013ApJ...765..131K,2014ApJ...787L..29K}. 
The core assumption of RCEs is that the atmosphere can be represented by a single vertical column. The atmospheric structure (i.e., pressure, temperature, and composition profiles) is first computed, typically assuming a pseudo-adiabatic lapse rate \citep{1988Icar...74..472K,2010ppc..book.....P}, where condensed species are immediately removed via precipitation. Radiative transfer is then calculated by estimating the stellar flux absorbed (including Rayleigh scattering) and the planet’s thermal emission (see Fig.~\ref{fig:Fig_tools}).

Initially, radiative equilibrium is not guaranteed. 
To achieve it, the model either adjusts the surface temperature (referred to as inverse modeling) or iteratively updates the atmospheric temperature profile based on computed heating rates in each layer, known as time-marching RCEs. 
Inverse methods converge quickly, while time-marching approaches can be computationally intensive, especially for hot, moist atmospheres with large thermal inertia, where numerical acceleration techniques may be required \citep[{as we used for}][]{2023Natur.620..287S}.



\subsubsection{1-D Energy Balance Models}\label{subsub:1DEBM}

1-D latitudinal Energy Balance Models (EBMs) do not calculate the radiative transfer like in RCEs, are latitudinally resolved (instead of vertically), allowing them to partially account for horizontal heat redistribution in a planet (see Fig.~\ref{fig:Fig_tools}). Their fundamental principle is radiative balance, essentially to calculate the balance between incoming energy from an external source (e.g., the Sun) and outgoing energy from a reservoir (e.g., a planet). 
They are widely used to study planetary climates, including those of Earth and exoplanets \citep[e.g.,][]{spiegel2008, dressing2010, 2024GeoRL..5109512C}. 
Their reduced complexity allows for rapid simulations and broad parameter space exploration, making them particularly useful for studying long-term climate evolution and sensitivity to key parameters. 
Despite their simplicity, they capture essential climate processes and by doing so, EBMs provide valuable insights at a very low computation cost, thereby complementing more detailed GCM studies.


The independent seminal works of \citet{budyko1969} and \citet{sellers1969} demonstrated the robustness of 1-D EBMs in studying Earth's present-day "warm state," characterised by high-latitude ice caps. Notably, \cite{budyko1969} also highlighted the existence of a critical ice-coverage threshold, beyond which the Earth could transition into a globally glaciated "Snowball" state - an early example of the sensitivity of Earth's climate to the ice-albedo effect.

A 1-D latitudinal EBM has been in development at the University of Geneva (\citealt{bhatnagar_ebm}, \url{https://github.com/DynaClim/ebm/tree/main}), building upon the frameworks of \cite{williams&kasting1997}, \cite{spiegel2008}, and \cite{dressing2010}. Written in \textit{Rust}, the model achieves high computational efficiency while offering multiple numerical integrators to balance accuracy and performance.
Energy redistribution is parameterised as temperature-driven diffusion, which can be linked to the planetary rotation rate \citep[see][]{williams&kasting1997, dressing2010}.
The model allows for flexible surface configurations (aquaplanet, land planet, or one featuring both continents and oceans) and albedo values that vary as a function of temperature and stellar spectrum \citep{spiegel2008, dressing2010, gilmore2014}.
Orbital parameters such as eccentricity, obliquity, orbital distance, and initial position can also be changed. The model has already been used to simulate exotic climatic scenarios (see Section~\ref{sub:Earth_habitability} for more details).

\subsubsection{3-D Global Climate Models}\label{subsub:3D_models}

3-D Global Climate Models (GCMs) are among the most sophisticated tools used to study planetary habitability. 
Whether developed from scratch \citep[e.g., THOR{, a model developed within PlanetS};][]{2020ApJS..248...30D} or adapted from Earth-based models \citep[e.g., LMDZ;][]{2006ClDy...27..787H}, all GCMs are built upon two core physical components: (1) the solution of the primitive equations of geophysical fluid dynamics, and (2) radiative transfer through the atmosphere. Models vary in how they treat these components—employing different numerical methods, radiative schemes, and physical parameterizations—and may incorporate additional processes such as moist convection, cloud microphysics, precipitation, photochemistry, and surface interactions (e.g., ocean and ice dynamics).{ However, as mentioned earlier, the increased physical fidelity of GCMs comes at the cost of many more input parameters and higher computational demands than 1-D models.}

THOR is the first open-source GCM developed from the ground up specifically for exoplanetary studies, designed to avoid common Earth-centric assumptions \citep{2016ApJ...829..115M}. It solves the full non-hydrostatic Euler equations on an icosahedral grid, avoiding approximations such as hydrostatic equilibrium. Recent upgrades \citep{2020ApJS..248...30D} have improved its numerical stability, expanded its physical modules, and enhanced its capacity to simulate diverse planetary atmospheres \citep[e.g.,][]{2020A&A...638A..77A,2022MNRAS.512.3759D,2022A&A...663A..79A,2023MNRAS.524.3396N}.


The Generic Planetary Climate Model (Generic-PCM) \citep{forget_inprep} originated at the Laboratoire de Météorologie Dynamique (LMD) in Paris and is now developed by a broader community, including {PlanetS through} the University of Geneva. Initially designed for Earth's climate, the model was progressively adapted to simulate Mars \citep{Forget:1999}, early Mars \citep{Wordsworth:2013}, and a wide range of exoplanets. This stepwise approach enabled the model to be validated under extreme conditions—from Venus’s hot, dense atmosphere to Pluto’s cold, tenuous one. Today, {we use} the Generic-PCM to study a wide diversity of climates focusing on cold rocky worlds \citep{2018A&A...612A..86T,Turbet:2022,2023A&A...680A.103C}. {However, the model is also used to study} hot gas giants \citep[e.g.][]{2021A&A...646A.171C,Teinturier:2024}. 
Notably, it is the only GCM currently capable of treating water vapor as a major atmospheric component, making it crucial for studying water-rich planets. {We also recently implemented} a {new} "dynamical slab ocean" module \citep{bhatnagar_ocean} as opposed to a static ocean (more in Sec.~\ref{subsub:exo-oceans}), which enables coupled atmosphere-ocean simulations with key processes such as sea ice/snow evolution, wind-driven Ekman transport \citep{codron2012}, horizontal diffusion, and convective adjustment.

\subsubsection{Atmospheric retrievals}\label{subsub:retrievals}

Atmospheric retrievals convert noisy spectral observations into estimates of planetary and atmospheric parameters. 
This process involves two key components - a forward model and a parameter inference algorithm. 
The forward model generates atmospheric spectra based on given planetary and atmospheric parameters. 
The parameter inference algorithm searches the prior space, defined by chosen prior probability distributions for model parameters, to find parameter combinations that best fit the input noisy spectrum \citep[e.g. {the PlanetS work of}][]{2017AJ....154...91L}. 
{We showed that} sometimes, machine learning techniques can also be used to find the best parameter combinations \citep[e.g.][]{2020AJ....159..192F}.
The main retrieval output is the posterior probability distribution, which contains the probability associated with any possible combination of model parameter values contained within the prior space. 
{We showed in PlanetS that }retrievals can be applied to any type of observations, such as phase curves \citep{
2021NatAs...5.1001H}, spectra from directly imaged planets \citep[e.g.][]{2017AJ....154...91L}, or ground based high-resolution spectra \citep{2020AJ....159..192F}.
More details can be found in the Chapter XXX by Kitzmann et al. 

Atmospheric retrievals also play a crucial role in context of the Large Interferometer for Exoplanets (LIFE) mission, a European-led initiative focused on the detection and characterization of temperate terrestrial atmospheres (see Section \ref{subsub:how_exo}). 
These retrievals serve dual purposes: assessing the observability of potential signatures of habitability and biospheres in emission spectra with simulated LIFE noise and with this refining mission requirements. 
The {PlanetS}-sponsored LIFE atmospheric retrieval routine, initially introduced in \citet{2022A&A...664A..23K} and subsequently enhanced by \citet{2022A&A...665A.106A} and \citet{2023A&A...673A..94K,2024ApJ...975...13K}, has been extensively utilized across various LIFE science cases over the last years (see Section \ref{subsub:how_exo}). 
Within the forward model of the retrieval routine, we use petitRADTRANS to calculate theoretical 1-D mid-infrared emission spectra. 
As parameter inference tool pyMultiNest \citep{Buchner_2016}, a Python interface for the MultiNest nested sampling package \citep{Feroz_2009}, is employed. The LIFE atmospheric retrieval routine has been recently released open-source for public use\footnote{\href{https://github.com/LIFE-SpaceMission/LIFE-Retrieval-Framework}{https://github.com/LIFE-SpaceMission/LIFE-Retrieval-Framework}}.

\subsection{Habitability in the solar system}\label{sub:Habitability_SS}





Since habitability corresponds to environments that can support life, and therefore, liquid water, the definition can extend beyond the classical habitable zone. 
Based on examples from our solar system, four distinct habitat classes have been proposed (see \cite{lammer2009,forget2013}):
\begin{enumerate}
    \item \textbf{Class I} habitats are bodies that can (currently) retain liquid water on their surface (Earth).
    \item \textbf{Class II} habitats refer to bodies that had Class I properties at some point in the past, but lost the ability to maintain liquid water at their surface (Mars).
    \item \textbf{Class III} habitats imply bodies that have a sub-surface liquid water ocean above a rocky core. The surfaces of such planets could either be rocky or icy (Europa, Enceladus).
    \item \textbf{Class IV} habitats refer to bodies that have a sub-surface liquid water ocean sandwiched between two layers of ice (Ganymede).
    \item \textbf{Class V} habitats are bodies — rocky or gaseous — that have clouds composed of suspended liquid water droplets in their atmosphere (Earth, and possibly Venus or Jupiter). 
\end{enumerate}
   
The following subsections will explore the PlanetS-related research pertaining to these classes of habitability.



\subsubsection{Earth}\label{sub:Earth_habitability}

Earth remains the only planet known to host life, making it the archetype of a habitable world (Class I habitat). 
It is the benchmark for assessing planetary habitability, both within the solar system and for exoplanets. Unlike other planets in our system, the Earth has sustained continuous surface liquid water for nearly 4 billion years\footnote{With the exception of brief global glaciation events around 2.45–2.22 Gya and 710–650 Mya, which were geologically short-lived \citep{evans1997,pierrehumbert2011}}. This remarkable feature is the result of a prolonged delicate balance between astronomical, geological, atmospheric and biological processes that regulate climate stability.

{ In \citet{2021Natur.598..276T}, we give} a compelling example of how astronomical and atmospheric processes work in tandem to maintain Earth's long-term climate stability. Using the 3-D GCM, the Generic-PCM, {we} demonstrated that Earth's current insolation (about 340.5 W m$^{-2}$) exceeds the water condensation insolation threshold (approximately 312.5 W m$^{-2}$), leading to two key implications. First, present-day Earth can exist in three main climate states: (1) its current state, with stable surface liquid water; (2) a snowball Earth, where oceans are nearly or entirely frozen; and (3) a steam Earth, where water exists predominantly as steam. Second, 4 billion years ago, when the Sun's luminosity was 75\% of its present value, the lower solar flux allowed water to condense and form stable oceans, setting the stage for the emergence and evolution of life. This suggests that the faint young Sun paradox, historically seen as a problem in explaining Earth's sustained surface liquid water, was actually crucial in establishing Earth's habitability.

A fundamental aspect of Earth's long-term climate stability, and by extension, its habitability, is its oceans. While oceans formed the stage for life's emergence, they also play a critical role in regulating Earth's climate through their immense heat capacity, energy transport across latitudes, and interactions with atmospheric circulation. 
Using the dynamical slab ocean model coupled to the atmosphere model of the Generic-PCM (see Section~\ref{subsub:3D_models}), {we} were able to reproduce key features of present-day Earth \citep{bhatnagar_ocean}, comparable to fully dynamic GCM studies \citep{marshall2007}. These include the major oceanic heat flows, an annual average surface temperature of 13$^\circ$C, planetary albedo of 0.32 and 20 million sq. km of sea-ice. This work highlights how Earth's oceans provide both a template for exoplanetary habitability and a testbed for improving climate models.

Using a version of the 1-D Energy Balance Model \citep{bhatnagar_ebm} introduced in Section~\ref{subsub:1DEBM}, {we} investigated Snowball Earth's response to stochastic events such as asteroid impacts and supervolcanic eruptions \citep{2024GeoRL..5109512C}. This study showed that even large impactors (diameter $\approx$ 100 km) and Toba-like eruptions (74 Kyr ago), are insufficient to deglaciate the Snowball state, unless atmospheric CO$_2$ had already been elevated through prolonged outgassing. This highlights the resilience of Snowball-Earth-like events, while also demonstrating the model’s versatility in simulating a wide range of climate scenarios.




\subsubsection{Early and modern Mars}\label{sub:Mars_habitability}


It is now firmly established that the climate of early Mars was very different from the one we know today. Valley systems and outflow channels \citep{Sharp75,Milton73} are large morphological remnants of a much wetter early-Mars (which makes it a Class II habitat). The presence of clay and sulfate at the surface, are also a mineralogical clue for water alteration. These have been observed and analyzed for decades, first from orbit \citep{Christensen00,Christensen01,Poulet07} and now in situ at the surface \citep{Vasavada22}.

However, most detected clays are Fe-Mg rich, and on Earth at least, this is more commonly associated with hydrothermal systems or groundwater alteration rather than sustained rainfall and large bodies of standing water \citep{Mangold07}. 
The overall detection of clays is also patchy, and localized, not homogeneous and widespread. Finally, from a modeling point of view, it is also difficult to find a thermal equilibrium in which a stable wet warm early Mars could co-exist with a faint, young Sun (but not impossible, e.g. \citealt{2021arXiv210310301T}). It is then likely that Mars had transient habitable environments, scattered across the (sub-)surface, rather than widespread and stable habitable conditions. Since then, the early climate of Mars has been and still is an ongoing debate in the community. \\
The surface of present-day Mars is an inhospitable environment for life and has likely been so for at least the past 3 billion years. The surface and the atmosphere are dry, cold and sterilized by Solar ultraviolet radiation. The first successful Mars landers (Viking 1 and 2) were equipped with various instruments to search for putative current life forms at the Martian surface \citep{Soffen76}. While the mass spectrometer did not detect any carbon, one experiment aiming at testing possible metabolic activity returned intriguing results \citep{Klein76}. It was only realized much later that the oxidative environment of the current Martian surface and in particular the presence of perchlorates in the surface material was responsible for the reactivity observed \citep{Quinn13}. 


Potential aquifers deep in the Martian sub-surface might be the only place where life could still exist, as well as in radiative habitable pockets within snow and ice \citep{Khuller2024}. Sounding radar observations of the southern polar layer deposits show a strong reflection potentially consistent with liquid water \citep{Arnold2022} but this interpretation is disputed \citep{Schroeder21}. Analyses of seismic data measured by InSight indicate the possibility of liquid water in the crust \citep{2024PNAS..12109983W}. Some Martian meteorites also show signs of aqueous alteration. 

More recently, detection claims of methane in the Martian atmosphere, first from telescopic observations and then in-situ by NASA's rover Curiosity have revived hopes for a present-day biological or volcanic activity at Mars. However, simultaneous in-situ and orbital observations provide inconclusive results and might indicate that the methane observed today diffuses slowly from sub-surface reservoirs formed much earlier in Martin history by unknown processes.

Due to its immensely rich past climatic history, Mars remains the best study location to one day, maybe bridge the gap between habitability studies, and astrobiology, by finding the first, unambiguous biosignature \citep{Hurowitz24}. 
Throughout its existence, the NCCR PlanetS has supported many studies to understand Mars better. Among them, the development and calibration of the ExoMars-TGO CaSSIS (\citep{Thomas2017_CaSSIS})instrument (\citep{Pommerol2022_CaSSIS_calib}) represent an important effort. The CaSSIS instrument can image the surface of Mars at 4.5m/px resolution in 4 broadband filters (BLU, PAN,RED, NIR) at different local times and viewing angles. The acquisition of images and the operation pipeline is now largely automated and operated from the University of Bern (\citep{Almeida2023_Targeting}). 
These capacities enabled scientists within the NCCR to contribute to the effort of answering key questions about Mars, from its surface to its atmosphere. Notable successes include the verification of the Kieffer model (\citep{Kieffer2007_ColdJets}) by documenting evolving spots’ morphologies through CaSSIS image series (\citep{Cesar2022_SouthPolarSpots}). Using CaSSIS’s multi-angular abilities (\citep{Valantinas2021_SlopeStreaks}) highlighted compositional differences between dark and bright slope streaks. Recently, observations of Mars limb revealed the vertical distribution of aerosols at the best resolution ever reached (\citep{Thomas2025_Millefeuille}). The NCCR also supports laboratory studies that enable the investigation of  Martian surface processes. Using facilities in the University of Bern Icelab (\citep{Pommerol2019_IceLab}), researchers investigate the relationship between frost formation and deposition on Mars regolith simulants(\citep{Spadaccia2023_FrostDetectability} \citep{Cesar2020_MGS1}). This allows to better understand remote sensing observations of Mars’ volatile cycle, a key component of the planet’s Modern climate (\citep{Valantinas2024_MorningFrost}). 
These results are detailed in the dedicated Mars chapter (Pommerol et al., this issue). 

\subsubsection{Venus}\label{sub:Venus_habitability}


The question of whether early Venus had liquid water oceans has been a long-standing debate in planetary science. Venus is now a scorching, dry world with a very thick CO$_2$-dominated atmosphere, but some studies have suggested it might have once had oceans. \citet{Way:2016,2020JGRE..12506276W} showed using the ROCKE-3D GCM \citep{way2017resolving} that early Venus could have had stable surface water for billions of years (which would make it a Class II habitat), provided it had a slow enough rotation and the right atmospheric conditions. 
Their models suggested that, with a faint young Sun, Venus could have maintained a temperate climate long enough for oceans to exist, due to a very efficient umbrella effect produced by water clouds accumulating on the dayside \citep{2013ApJ...771L..45Y,2014ApJ...787L...2Y}.
{In \citet{2021Natur.598..276T}, we} later challenged this view by using another GCM (the Generic PCM) to simulate the conditions of ocean formation at the end of the magma ocean, an inevitable step in the evolution of telluric planets. The simulations show that, on early Venus, water clouds would have preferentially formed on the nightside due to intense absorption of sunlight by water vapor on the dayside. These nightside clouds have a net warming effect, preventing the planet from cooling enough for water to condense into oceans, even at relatively low solar radiation levels. This suggests that Venus never had liquid water and that it followed a fundamentally different evolutionary path from Earth. {In \citet{Turbet:2023}, we} eventually showed that many close-in exoplanets are likely to have suffered a similar fate.

Venus' current dense atmosphere, rich in CO$_2$, is enveloped by cloud layers primarily composed of sulfuric acid droplets at altitudes ranging from 47 to 70 km \citep{delitsky_cloud_2023}. These clouds maintain temperatures that could theoretically support some extremophiles found on Earth \citep{cockell_life_1999, morowitz_life_1967} (which could make Venus a Class V habitat). One of the most debated recent discoveries is the detection claim of phosphine (PH$_3$), a biosignature gas, in Venus' atmosphere \citep{greaves_phosphine_2021, seager_venusian_2021, bains_phosphine_2024}, though its presence remains speculative \citep{2020A&A...643L...5E,2020A&A...644L...2S,2021NatAs...5..631V,2021ApJ...908L..44L}. This discovery has reignited interest in Venus' potential habitability and led to the proposal of dedicated missions to Venus, specifically designed to measure habitability indicators and search for life-related molecular building blocks within its cloud layers by a probe {by us, as part of the Morning Star Mission program, previously known as Venus Life Finder \citep{2022Aeros...9..312L, MSM_VLF_Venus}}.
This program will complement the planned missions to Venus (EnVision, \citealt{2012ExA....33..337G}; DAVINCI, \citealt{2022PSJ.....3..117G}; VERITAS, \citealt{2020LPI....51.1449S}) which will shed light on the potential past habitability of the planet.





\subsubsection{Ocean worlds and moons}\label{sub:Moons_habitability}

In the outer Solar System, many satellites are found to harbour sub-surface water oceans hidden below an icy crust. 
On Earth, extremophilic microbes thrive in deep-sea hydrothermal vents \citep[e.g.,][]{baross1985, martin2008}. Similarly, chemosynthetic life could emerge in sub-surface ocean worlds, where radiogenic decay and tidal dissipation can act as an energy source \citep{mckay2008}. 
This makes ocean worlds intriguing astrobiological targets. The most promising ocean worlds to search for signatures of life include Jupiter's moon Europa and Saturn's satellite Enceladus (both Class III habitats). Furthermore, there is evidence that Jupiter's moons Ganymede and Callisto, Saturn's natural satellite Dione, and Neptune's Triton harbour a sub-surface water ocean (Class IV habitats). Smaller moons of Saturn (e.g., Mimas, Tethys, Rhea, Iapetus) and Uranus (e.g., Miranda, Ariel, Umbriel, Titania, Oberon) are also ocean world candidates \citep{national_academies_of_sciences_engineering_and_medicine_origins_2022}.

For Enceladus, the Cassini mission measured an active water vapour plume system, composed of multiple jets \citep{porco2006, hansen2006}, which enables sampling the composition of the sub-surface ocean. 
Measurements with the Ion and Neutral Gas Mass Spectrometer (INMS) showed that Enceladus' plumes consist primarily of water. In addition, a range of volatiles \citep[e.g., CO$_2$, CH$_4$, NH$_3$, H$_2$;][]{waite2006} could be identified in the gas phase of the plumes. The Cosmic Dust Analyzer (CDA), on the other hand, measured solid particles entrained within the plumes (and Saturn's E-ring, which is formed by ice grains escaping Enceladus's plume into orbits around Saturn). The measurements not only revealed a class of salt-rich ice grains, providing information on a range of major solutes in the ocean water (Na$^+$, K$^+$, Cl$^-$, HCO$_3$$^-$, CO$_3$$^{2-}$) but also the presence of phosphorus, the least abundant of the bio-essential elements \citep{postberg2023}. Together, the measurements of both mass spectrometers (gas and solid) indicate that the plumes encountered on Enceladus are directly fed from the sub-surface ocean and that the ocean is in direct contact with a rocky seafloor.

Europa's induced magnetic field was discovered by the Galileo mission \citep[see][]{khurana1998, kivelson2000} and indicates that the moon has a sub-surface ocean. This ocean is also likely in contact with the silicate mantle \citep[see e.g.,][]{anderson1998}. Furthermore, there has been some evidence for active plumes on Europa \citep[e.g.,][]{roth2014, sparks2016, jia2018}, though due to their very indirect nature and uncertainties, as well as a very extensive list of unsuccessful follow-up searches \citep[e.g. most recently,][]{villanueva2023, hansen2024}, these potential plume detections are still under debate. ESA's Jupiter Icy Moons Explorer (JUICE) and NASA's Europa Clipper mission are both underway to explore Europa from up close. While both missions are equipped with an extensive list of instruments designated to analyze Europa's habitability (e.g., with the mass spectrometers NIM \citep{Foehn2021} on JUICE and MASPEX \citep{Waite2024} on Europa Clipper, as well as the dust detector SUDA \citep{Kempf2025} on Europa Clipper), JUICE will only have two flybys of Europa at an altitude of approximately 400~km, while Europa Clipper will fly by Europa almost 50 times with closes approaches as low as 25~km. Both Europa Clipper and JUICE aim to assess Europa's habitability by studying its ice shell, sub-surface water, surface composition, and geological activity. Together, these missions will provide a comprehensive understanding of Europa's potential to support life by combining in-depth surface and sub-surface analysis with compositional and geological insights.

There is also the possibility that these habitats have more "exotic" life. For example, methane, ammonia, or sulfuric acid could act as a solvent to facilitate molecular interactions instead of water \citep[e.g.,][]{mckay2005, bains2021, bains2024}. Titan, Saturn's largest moon, has significant habitability potential due to its complex chemistry, sub-surface ocean, and Earth-like atmospheric processes \citep[e.g.,][]{MacKenzie2021}. Its thick nitrogen-rich atmosphere, laced with organic molecules, creates a dynamic system where pre-biotic chemistry could occur \citep[e.g.,][]{McKay2016}. Methane and ethane rivers, lakes, and seas on the surface suggest active hydrological cycles, albeit with liquid hydrocarbons instead of water. Beneath its icy crust, a global sub-surface ocean of liquid water, possibly mixed with ammonia, provides a potential environment for microbial life. Titan's energy sources, such as tidal heating and chemical disequilibrium, could drive metabolic processes, making it an intriguing target in the search for extraterrestrial habitability. NASA's Dragonfly mission will assess Titan's habitability by exploring its surface chemistry, atmospheric processes, and potential sub-surface ocean \citep{Turtle2024, Barnes2021}. By sampling and analyzing organic-rich materials, Dragonfly will investigate pre-biotic chemistry and search for complex molecules that could support life. Its mobility will allow it to study multiple locations, providing a broader understanding of Titan's diverse environments and their potential for sustaining life.

ESA is currently refining its next large-class mission, L4, as part of the Voyage 2050 program, focusing on the moons of the giant planets \citep{martins2024}. The mission aims to investigate habitability, pre-biotic chemistry, and potential biosignatures. To support the mission's initial definition, ESA assembled an expert committee to assess potential targets. After evaluating various options, the committee identified Saturn's moon Enceladus as the most compelling destination. They recommended a mission featuring in-situ sample acquisition, possibly through a lander or plume flythroughs to access sub-surface material. The proposed mission architecture includes a south polar lander, an orbiter, and a plume sampling system, enabled by a dual launch configuration. The committee also highlighted Titan and Jupiter's moon Europa as important candidates for future exploration.

{PlanetS contributions to characterise ocean worlds included the development of the deep-learning powered tool \textit{LineaMapper} \citep{haslebacher2024, haslebacher2025} to map surface features on Europa. Furthermore, laboratory experiments of surface ice analogues were performed for Europa \citep[e.g.,][]{ottersberg2025}.}. More details can be found in Chapter XXX by Haslebacher et al. 


\subsection{Habitability of exoplanets}\label{sub:Habitability_Exo}

The solar system serves as a vital benchmarking tool for exoplanet studies. Since exoplanets remain distant and only partially characterized, it is crucial to validate our models to study them against well-observed Solar System bodies. 
Observations and modelling of Venus, Earth, and Mars provide critical reference points for interpreting the climates of terrestrial exoplanets, while icy moons demonstrate the potential for life in non-traditional environments. 
By studying these diverse environments in our own system, we improve our ability to assess exoplanetary habitability.



\subsubsection{Concept of the habitable zone}\label{sub:HZ}



\paragraph{\textbf{The inner edge of the HZ}}\label{subsub:inner_HZ}

Thanks to fast computation times, the radiative transfer of 1-D RCE models can be performed by calculating state-of-the-art absorption spectra in every vertical layer of the atmosphere (i.e. line-by-line calculation). This allows to have an exact calculation of the absorbed and emitted fluxes for benchmarking radiative transfer calculation of sensitive science cases such as the onset of the runaway greenhouse. 
This allowed {us in }\citet{2022A&A...658A..40C} to solve the open question of a possible overshoot of the runaway greenhouse thermal emission limit (named the Simpson-Nakajima limit). 
{We} showed that, for a H$_2$O+N$_2$  atmosphere, such an overshoot is induced by the pressure broadening of H$_2$O absorption lines, due to interspecies collisions. 

The runaway greenhouse outcome has been explored using 1-D models as a possible evolutionary pathway for water-rich terrestrial planets and as a potential explanation for the divergence between Earth and Venus \citep{1988Icar...74..472K, 2021ApJ...919..130B}. {We showed in \citet{2019A&A...628A..12T,2020A&A...638A..41T} that} the radius inflation induced by the entire evaporation of deep water oceans could cause a significant change of the visible planetary radius. This radius gap between habitable and post-runaway planets could be detected by the PLATO mission \citep{2024PSJ.....5....3S}. 
{We also performed }another improvement of the mass-radius relationship for steam atmosphere planets {in \citet{2021ApJ...922L...4D} where we} accounted for the mixing of water in the surface magma ocean (which happens for sufficiently high surface temperatures).
More recently, {we showed in} \citet{2023Natur.620..287S} showed that assuming radiative-convective equilibrium profiles—valid for thin atmospheres like Earth’s—introduces large errors for modeling water-rich post-runaway atmospheres, overestimating surface temperatures by several hundred Kelvin. This result marks a paradigm shift in 1-D modeling of post-runaway planets and carries important implications for interpreting their mass-radius relationships. 

\begin{figure}[ht!]
     \centering
     \includegraphics[width=1\linewidth]{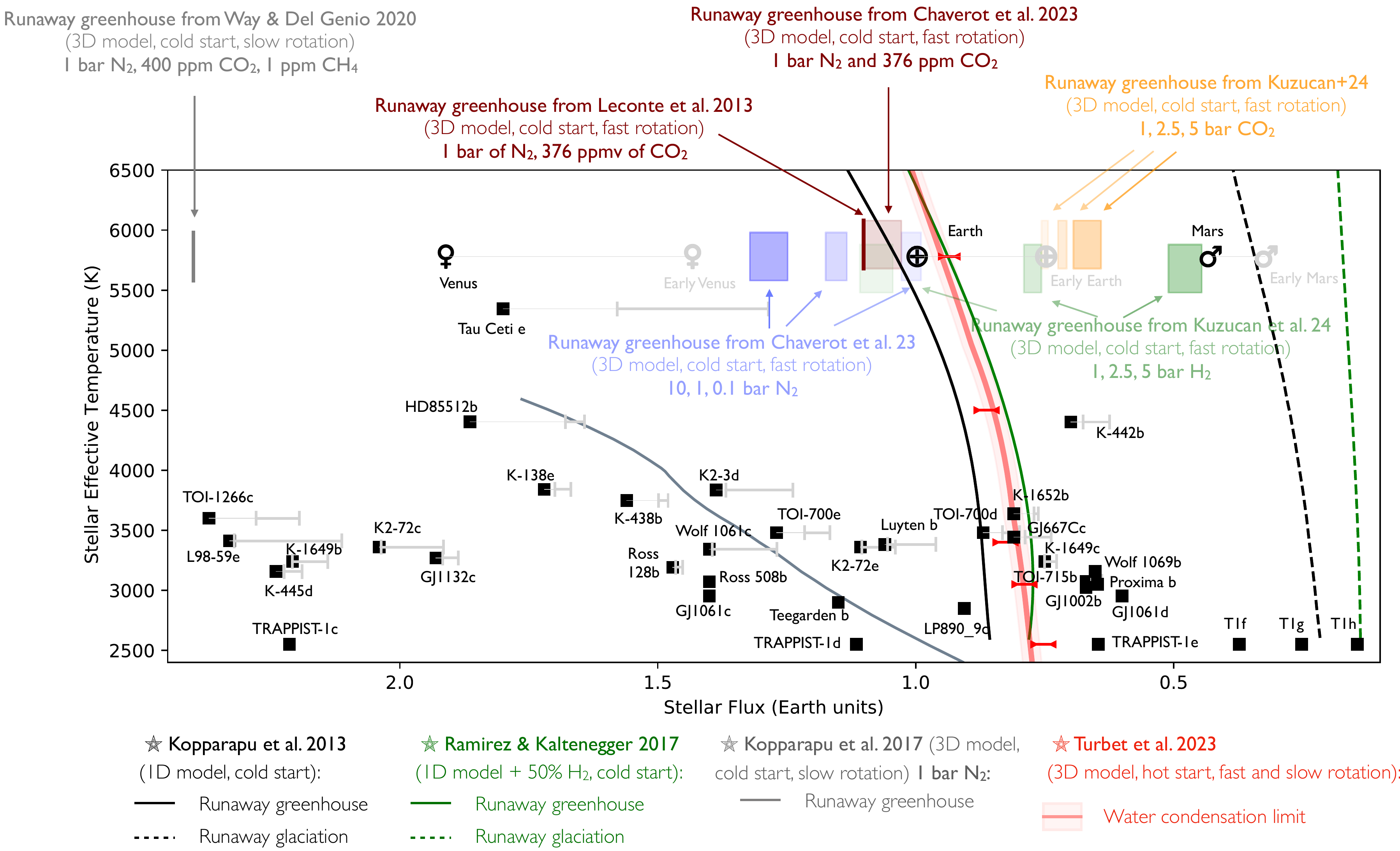}
     \caption{Different limits of the Habitable Zone for different (1-D and 3-D) models, different assumptions for the spin of the planet, composition of the atmosphere and its pressure. Most inner edge limits are given for the runaway greenhouse transition, but the red limit is given for the water condensation limit of \citep{Turbet:2023}. This figure is inspired from Figure~8 of \citep{Turbet:2023} with additions from more recent works like \citep{2023A&A...680A.103C} and \citep{2025Life...15...79K}.}
     \label{fig:HZ}
 \end{figure}

Recent studies using the Generic-PCM (see Sec.~\ref{subsub:3D_models}) have provided new insights into the climate evolution of rocky planets and the conditions required for surface water to condense or persist. 
These investigations highlight the importance of 3-D processes, such as atmospheric circulation and cloud dynamics, in shaping planetary climates-factors that are often missing in 1-D models.
{In }one key study{, we explored} the runaway greenhouse transition \citep{2023A&A...680A.103C}. 
By modeling this transition in 3-D, {we showed} how evaporation enriches the atmosphere with water vapor, followed by a rapid increase in surface temperature once the ocean is fully vaporized. 
The results emphasize the critical role of cloud coverage and atmospheric dynamics in this process and suggest that the runaway greenhouse, once triggered, is difficult to reverse due to radiative imbalances and a change in the behavior of the clouds.
{In }\citet{2022A&A...658A..40C} and \citet{2023A&A...680A.103C} (1-D and 3-D modeling, respectively){, we} also highlighted the influence of the nature and pressure of the background gas on the runaway greenhouse transition. 
For instance, increasing the pressure of N$_2$ in the atmosphere allows to push the inner boundary of the HZ closer to the star, and adding a little CO$_2$ in an otherwise pure N$_2$ atmosphere has the opposite effect.
Additionally{, we showed in} \citet{2025Life...15...79K} that increasing the pressure of a background gas which is also a greenhouse gas (pure CO$_2$ and pure H$_2$ atmosphere) has the impact of pushing the HZ inner boundary farther away from the star.
These studies have contributed significantly to the knowledge that there is not one HZ: its boundaries are highly dependent on atmospheric composition and pressure. 
Figure~\ref{fig:HZ} shows these results in a broader context.

Once the runaway greenhouse phase is complete and all surface liquid water has evaporated, the planet is left with a steam-dominated atmosphere—potentially representative of the early atmospheres of young, water-rich planets.
Building on the discoveries of \citet{2021Natur.598..276T} (see Sec.~\ref{sub:Earth_habitability} and \ref{sub:Venus_habitability}), {we explored in} \citet{Turbet:2023} the conditions necessary for ocean formation on rocky planets. It revealed a key asymmetry: the stellar flux required to condense a planet’s primordial water reservoir is significantly lower than the flux needed to vaporize an existing ocean. This leads to the definition of a new ``water condensation limit'', which lies well interior to the traditional inner edge of the habitable zone. The study also investigates observational signatures of steam-rich atmospheres on exoplanets, using TRAPPIST-1b, c, and d as case studies, and demonstrates that JWST may be able to detect such atmospheres and even constrain the presence of nightside water clouds.

\paragraph{\textbf{The outer edge of the HZ}}\label{subsub:outer_HZ}

%
%

The boundaries of the HZ are influenced by multiple factors, including atmospheric composition, orbital dynamics, and stellar type. At the outer edge, CO$_2$ condensation can trigger atmospheric collapse \citep{2020A&A...638A..77A} (see also Sec~\ref{subsub:collapse}), with silicate weathering also playing a role in regulating CO$_2$ levels \citep{2021PSJ.....2...49H}. Additionally, different atmospheric compositions can redefine habitability limits. For instance, {we showed that }H$_2$-rich atmospheres can extend the HZ \citep{2022NatAs...6..819M}, highlighting the complexity of planetary climate regulation across diverse stellar environments.

{In \cite{Hansen_2025}, we} evaluated {the ability of the LIFE mission (see Section \ref{paragraph:LIFE} for a detailed description of the LIFE mission concept and related works)} to detect $\mathrm{CO_{2}}$ trends indicative of the carbonate-silicate weathering feedback using thermal emission spectra representing atmospheric variability in the HZ. This concept, when applied to HZ planets at different orbital distances, predicts a decreasing trend in atmospheric $\mathrm{CO_{2}}$ with increasing incident stellar flux \citep{1993Icar..101..108K}. Detecting this trend provides a unique observable linked to the long-term presence of liquid surface water, a prerequisite for the carbonate-silicate cycle \citep{Walker_1981}. The study found robust detection of population-level $\mathrm{CO_{2}}$ trends for as few as 30 Exo-Earth Candidates (EECs), even at the lowest assessed spectrum quality (signal-to-noise ratio S/N = 10, spectral resolution R = 50). These findings indicate that $\mathrm{CO_{2}}$ trends represent a viable signature for detecting exoplanet habitability. {As a population-level tracer, such a trend is unable to directly assess the habitability of individual planets, where characterization may remain ambiguous, however, it provides a compelling comparative planetology framework to evaluate habitability across planetary populations.}

Finally, for planets orbiting cooler stars (M and K types), tidal locking and spin-orbit resonances significantly impact climate. Higher-order resonances, such as 3:2, lead to varying dayside conditions and equator-to-pole temperature gradients \citep[e.g.,][]{2016A&A...596A.112T, 2017A&A...601A.120B}. 
Eccentricity enhances these effects{: it increases} overall warming \citep{2016A&A...591A.106B} and {we showed that it} shifts the inner HZ boundary outward due to a wetter stratosphere \citep{2025PSJ.....6....5B}. Such resonances also introduce cyclic variations in UV exposure, temperature, and water abundance—key factors for habitability.

\subsubsection{Limitations of the HZ concept: accounting for evolution}\label{sub:limits_to_HZ}

\paragraph{\textbf{Change of composition: carbon cycle - all links with interior}}\label{subsub:interior}

If volcanic outgassing continues on Earth at the present-day rate, the partial pressure of atmospheric CO$_2$ ($\sim$0.4~mbar) would become 1 bar in 20 Myr \citep{Walker_1981}. 
This would rapidly increase the surface temperature due to the greenhouse effect of CO$_2$.
However, there is sufficient evidence that such a drastic change in the CO$_2$ partial pressure did not happen in the past and likely will not occur in the near future \citep{2001JGR...106.1373S,2018PNAS..115.4105K}. 
Moreover, the incident solar radiation (insolation) received by Earth has gradually increased over 4.5 billion years.
The current atmospheric CO$_2$ levels would not have been sufficient during the Archean to provide greenhouse warming to maintain a temperate climate.
This is also known as the Faint Young Sun Paradox \citep{sagan1972}. 
There is evidence that the CO$_2$ partial pressure was of the order of 0.1~bar during the Archean, likely just enough to prevent the Earth from freezing over \citep{catling2020}.

The carbonate-silicate cycle helps counterbalance the effects of volcanic outgassing on Myr timescales and the effects of an increase in solar luminosity on Gyr timescales \citep{1993Icar..101..108K,2024SpScT...4...75G}.  
The increase in atmospheric CO$_2$ produced by volcanism is attained by a process called silicate weathering \citep{urey1952,Walker_1981,berner1983}. 
Continental silicate weathering draws down atmospheric CO$_2$ by reactions with rainwater and silicate rocks \citep{Walker_1981}. 
The carbonate ions are then transported by rivers to oceans, where carbonates are precipitated, thus transforming silicates to carbonates \citep{kump2000}. 
This process of carbonate precipitation locks up atmospheric CO$_2$ in carbonate minerals on the seafloor \citep{zeebe2003}. 
Plate tectonics eventually subducts the seafloor carbonate into the upper mantle, thereby removing CO$_2$ from the atmosphere--ocean system \citep{2001JGR...106.1373S}. 
Although the near-surface CO$_2$ budget of Earth is similar to that of Venusian atmosphere, more than 99\% of near-surface CO$_2$ on Earth is expected to be locked up in the form of carbonates, courtesy of the carbonate-silicate cycle \citep[e.g. ][]{hartmann2012}.
Although it is important to note that plate tectonics may not be necessary for carbonate burial and recycling \citep[e.g., CO$_2$ cycling for stagnant-lid planets][]{foley2018,2019A&A...627A..48H}.
The high pressures and high temperatures of the upper mantle metamorphise carbonates back to silicates and release CO$_2$ via volcanism, and the carbonate-silicate cycle is closed.

On modern Earth, granitoid rocks containing felsic minerals are predominantly responsible for driving silicate weathering. 
However, the continental rocks in the past would have been more mafic than on present-day continents \citep{catling2020}. 
Even Earth-like exoplanets with volcanic islands instead of continents would mean a basaltic composition \citep{2015ApJ...812...36F}. 
The diversity of refractory elemental abundances in stellar photospheres (e.g., Mg/Fe, Mg/Si) suggests a much larger diversity of interior and surface composition of rocky exoplanets than can be visualised with the knowledge of the rocky bodies in the solar system \citep{2023ApJ...948...53S}. 
Thus, the carbonate-silicate cycle on rocky exoplanets should be sensitive to their composition. 
Silicate weathering is a strong function of rock composition as observed with field measurements and theoretical calculations \citep{2016GeCoA.190..265I,2018E&PSL.485..111W}. 
{We showed in }recent calculations that peridotite, an upper mantle rock, exhibits an order of magnitude higher weathering rate than granite present on Earth's continents \citep{2021PSJ.....2...49H}. 
Similarly, the stability field of carbonate precipitation is sensitive to the composition of carbonates \citep{2023ApJ...942L..20H}. 
Fe- and Ca-carbonates have a marginally wider stability field in parameter space defined by the surface temperature and the CO$_2$ partial pressure than Mg-carbonates. 
Thus, rocky exoplanets with a lack of Ca-carbonates but with the presence of Fe- or Mg-carbonates can still stabilise carbonate-silicate cycling. 
However, the coupled effect of the surface rock composition with subduction and outgassing remains to be evaluated.  

In addition to rock composition, the availability of weathering reactants, i.e., water, silicate rocks and CO$_2$ determines three limits of weathering \citep{kump2000,west2005,maher2014}. 
Kinetics of mineral dissolution reactions give rise to the kinetic limit, which is the case for modern Earth \citep{Walker_1981,kump2000}. 
If fresh silicate rocks cannot be supplied by plate tectonics, weathering becomes rock supply-limited \citep{west2005,2015ApJ...812...36F}.  
If there is limited availability of water, weathering becomes thermodynamically limited \citep{maher2014,2018E&PSL.485..111W,2021PSJ.....2...49H}. 
At the thermodynamic limit, weathering decreases as a function of temperature, resulting in a positive weathering feedback, as opposed to the negative feedback at the kinetic limit \citep{2021PSJ.....2...49H}. 
The impact of such positive feedback is not fully understood yet.

\paragraph{\textbf{Atmospheric escape}}\label{subsub:escape} 

One requirement for exoplanet habitability is the presence of an atmosphere: without it, there is no possibility to have surface liquid water.
This requirement is particularly important for exoplanet science as the atmosphere is the only window to probe the presence of {signs of} life on a planet (see next Section~\ref{sec:biosignatures}). 
However, an atmosphere can be lost to space, a process particularly efficient when the planet orbits an active star which emits high energy radiations.
Atmospheric escape processes are among the most important processes governing the evolution and the composition of the atmosphere of rocky planets \citep[e.g.][]{2022ARA&A..60..159W}

The processes contributing to the loss of the atmosphere can be classified in two groups: thermal processes and non-thermal processes \citep[see reviews from][]{2015AREPS..43..459T,2025RvMPP...9...18H}.
Traditionally, the exoplanet community has mainly considered thermal processes \citep[like Jeans escape or energy-limited escape, e.g.][]{2019AREPS..47...67O} while the solar system community has focused on non-thermal processes \citep[like pickup ion escape and sputtering, e.g.][]{2008SSRv..139..399L}. 
This comes from the fact that thermal processes are thought to be dominant for the very close-in planets for which escape has been observed \citep[e.g.][]{2004ApJ...604L..69V,2015Natur.522..459E}.
For less irradiated planets, such as planets in the Habitable Zone of low mass stars, both thermal \citep[e.g.][]{2015AsBio..15..119L,2017MNRAS.464.3728B} and non-thermal processes have been considered \citep[e.g.][]{2018PNAS..115..260D,2016ApJ...833L...4G,2022ApJ...941L...8G}.
For small rocky planets, the question of whether or not they can retain their atmosphere led to the concept of the cosmic shoreline now being empirically tested with the JWST: this limit separates the planets which have lost their atmosphere from the planet which managed to retain their atmosphere in the age of the system \citep{2017ApJ...843..122Z}.

We focus here on thermal escape processes and most precisely the energy-limited escape which is the most efficient thermal escape process.
Considering an energy-limited escape allows to derive upper values of the loss of the atmosphere and thus calculate a worst case scenario for the planets' potential habitability.
This was applied to the TRAPPIST-1 system \citep{2017Natur.542..456G}, in the specific context of water loss.

{In }\citet{2017AJ....154..121B}{, we} conducted a reconnaissance with the Hubble Space Telescope to investigate the Lyman-$\alpha$ emission of TRAPPIST-1, assess the presence of hydrogen exospheres around its two inner planets, and determine their UV irradiation levels. 
{We} detected the Lyman-$\alpha$ line of TRAPPIST-1, and showed that compared to Proxima Centauri \citep{2016Natur.536..437A,2016A&A...596A.111R,2017A&A...603A..58R}, which has a similar X-ray output, TRAPPIST-1 exhibits much lower Lyman-$\alpha$ emission. 
Using these findings and the method of \citet{2017MNRAS.464.3728B}, {we} estimated atmospheric mass loss rates for all the planets in the system \citep{2017A&A...599L...3B}, concluding that while TRAPPIST-1 has moderate extreme UV emission, the total XUV irradiation could still be strong enough to strip the atmospheres of the inner planets within a few billion years.
However, the Habitable Zone planets, like TRAPPIST-1e, might have lost less than three Earth oceans if hydrodynamic escape stopped once they entered the Habitable Zone.
These results were derived before the latest estimations of the masses of the planets \citep{2021PSJ.....2....1A}, so these numbers should be taken as orders of magnitude.
Since these studies, JWST has shown that the inner planets of TRAPPIST-1 might not have atmospheres \citep{2023Natur.618...39G,2023Natur.620..746Z} although there is still some hope for the outer planets \citep{2024ESS.....550002K}.

\paragraph{\textbf{Stability: atmospheric collapse}}\label{subsub:collapse} 

The stability of atmospheres on tidally-locked planets has been a long-standing question. These planets may exhibit a strong temperature gradient between their permanent night and daysides. The temperature of the nightside depends crucially on the efficiency with which heat can be redistributed from the illuminated day to the nightside. This, in turn, will determine the stability of the nightside atmosphere against a potential collapse due to the potentially low temperature and the freeze out of the major atmospheric constituents, such as carbon dioxide.

By using a hierarchy of analytic zero-dimensional and one-dimensional atmospheric models, {we} studied the atmospheric stability of tidally locked, terrestrial exoplanets, taking into account the coupling between radiative transfer, convection, and the atmosphere's large-scale circulation \citep{2020A&A...638A..77A}. The study explored the stability limits as a function of the atmospheric shortwave and longwave opacities and the total surface pressure. The study showed, in particular, that the atmospheric stability on the nightside decreases with increasing planet mass. This is caused by a decay of atmospheric optical depth, and thus of the greenhouse effect, due to the atmosphere of a heavier planets having a smaller vertical extent. Comparisons with more complex calculations with full global climate models showed the these simplified models can capture the most important effects and can, therefore, be used to infer the potential atmospheric collapse on the nightside.

\paragraph{\textbf{Oceanic heat transport}}\label{subsub:exo-oceans} 

The classical HZ framework primarily considers radiative balance and often neglects ocean dynamics. Additionally, most exoclimate models omit dynamic oceans due to their high computing cost. This oversight is significant, as ocean dynamics can critically impact the climate and observables of temperate exoplanets \citep{yang2019ocean}. As a compromise, most exoplanet GCMs use static (``slab'') ocean models without ocean dynamics. 
{In }\cite{bhatnagar_trappist}, {we} use an improved compromise: the dynamical slab ocean model \citep{bhatnagar_ocean} of the Generic-PCM - and apply it to TRAPPIST-1e. 
Previous studies have suggested that if TRAPPIST-1e had formed with a substantial water reservoir  \citep{tian2015water,2024ESS.....550002K}, it could have sustained liquid water oceans \citep{2018A&A...612A..86T}. 
Using a static slab ocean, TRAPPIST-1e exhibits the familiar ``eyeball'' climate with a hot dayside and frozen nightside. 
However, \cite{bhatnagar_trappist} find that with ocean heat transport enabled, the climate transitions to a more exotic structure reminiscent of the ``lobster'' pattern identified for Proxima Centauri b by \cite{2019AsBio..19...99D} using a fully dynamic ocean model. These patterns arise from interactions between atmospheric and oceanic Rossby and Kelvin waves under tight coupling. 
\cite{bhatnagar_trappist} thus demonstrate how ocean heat transport can significantly alter exoplanet climates—capturing key dynamical behavior with far lower computational cost than fully dynamic ocean models.

\section{Biosignatures and their detectability}\label{sec:biosignatures}


\subsection{What are we looking for?}\label{sub:what_look_for}


The concept of biosignatures is central to the search for extraterrestrial life, yet defining it comprehensively remains a challenge. Since the definition of life itself is still a subject of debate, biosignatures are inherently difficult to characterize. In practice, they are often identified based on traits observed in terrestrial life, which primarily relies on carbon-based chemistry and liquid water as a solvent \citep{westall2018}. However, some researchers argue that a universally definitive biosignature may not exist, as the term can be misleading and subject to varying interpretations \citep{malaterre2023}. The gradual transition from simple chemistry to fully developed life forms creates a ``grey area'' where distinguishing biological processes from purely abiotic ones becomes increasingly complex. This ambiguity highlights the need for a refined yet flexible framework for identifying biosignatures in planetary exploration.

NASA’s Astrobiology Roadmap defines a biosignature as ``an object, substance and/or pattern whose origin specifically requires a biological agent'' \citep{desmarais2008}. 
{However, this strict definition is not universally adopted. A large part of the astrobiology community considers biosignature gases as potential indicators of metabolism, even without specifying the underlying biology. 
In this view, such gases are treated as plausible by-products of life-sustaining processes, regardless of whether the detailed biochemistry resembles that of Earth.}
{In all cases}, the context in which a biosignature is studied is of importance, as the indicators of life vary between planetary environments within the Solar System and those found in exoplanetary atmospheres. Given these distinctions, this chapter will discuss biosignatures separately in the context of the Solar System and exoplanets.





\subsubsection{Solar system}\label{subsub:what_SS}



The search for biosignatures in the Solar System is greatly enhanced by observations from orbit, but also in-situ exploration missions, such as probes, rovers, and sample-return missions, which allow direct investigation of planetary atmospheres, surfaces and sub-surface environments.

Biosignatures can be categorized into four main groups based on their characteristics \citep{malaterre2023}. Substance-based biosignatures include chemical constituents such as amino acids, peptides, lipids, and DNA, which are essential components of life on Earth. The presence of these molecules, especially in specific ratios or distributions, can suggest biological activity, though some may also form abiotically \citep{callahan_carbonaceous_2011, glavin_chapter_2018}. Structural biosignatures encompass microstructures and macrostructures, as well as molecular complexity indicative of biological processes. For example, complex biopolymers like RNA are unlikely to arise without biological intervention \citep{neveu_ladder_2018}, while fossilized microbial formations or stromatolitic structures may provide evidence of past life. Process-based biosignatures involve observable biological mechanisms such as thermodynamic and redox disequilibrium, biogeochemical cycles (e.g., nitrogen cycles), and evolutionary markers consistent with Darwinian evolution. These processes create specific environmental signatures that can indicate life’s presence. Pattern-based biosignatures include isotopic variations, isotope fractionation, homochirality (e.g., the exclusive use of L-amino acids in terrestrial proteins), and lipid chain length distributions, all of which can provide insights into potential biological activity \citep{weller_mystery_2024}.

Both chemical and physical biosignatures are critical in astrobiology \citep{2017LPICo1989.8141H}. 
Physical biosignatures, such as microbial mats, microfossils, and atmospheric fluctuations influenced by biological processes, offer macroscopic evidence of life. Chemical biosignatures, including unique mineral compositions and specific molecular distributions, are equally important \citep{2017LPICo1989.8141H, mustard_report_2013}. Since Earth’s life is based on a common biochemical framework using CHNOPS elements \citep{national_academies_of_sciences_engineering_and_medicine_origins_2022}, identifying similar biochemical patterns elsewhere could provide strong evidence for extraterrestrial life. However, to address the possibility of non-Earth-like biochemistries, researchers are developing agnostic biosignature detection methods that focus on unexpected molecular complexity, chemical disequilibrium, and anomalous chemical distributions rather than specific known biological molecules \citep{bartlett_defining_2020, neveu_ladder_2018}.

Given the potential scarcity and uneven distribution of biosignatures on other planetary bodies, future space missions must employ highly sensitive instruments and strategic site selection to maximize the likelihood of detecting signs of life \citep{aerts_biota_2014, westall_biosignatures_2015, mackenzie_enceladus_2020}. By refining detection techniques and expanding the scope of biosignature classification, scientists can enhance the search for {signs of }life, whether it mirrors Earth’s biology or represents a completely novel form of life.


\subsubsection{Exoplanets}\label{subsub:what_exo}



Contrary to Solar System exploration, exoplanet research has been confined to the realm of remote observations. Therefore, the ``signatures" that would suggest the presence of life on an exoplanet must be observed in the spectrum of the exoplanet. 
We can categorize exoplanet-related biosignatures into three primary types: gaseous, surface, and temporal \citep[see, e.g.,][]{2018haex.bookE..68G, 2019Galax...7...82C,2024RvMG...90..465S}.

Gaseous biosignatures are the spectral fingerprint of life-related molecules in the atmosphere. Some examples are:
\begin{itemize}
    \item Molecular oxygen (O$_2$). Oxygen is produced by photosynthesis and can accumulate in the atmosphere if production outweighs sinks, as it happened on Earth during the Great Oxygenation Event and the Neoproterozoic Oxygenation Event \citep[see][and references therein]{2024RvMG...90..465S}. It can be observed in the visible portion of the spectrum, with very strong lines around 0.7 micron, and with dimmer features in the near infrared mostly related to collision-induced absorption.
    \item Ozone (O$_3$). Ozone is a photochemical byproduct of molecular oxygen. It is often used as an indirect tracer for O$_2$ because of the strong bands in the UV and MIR which are generally easier to detect. The relationship between ozone and oxygen is however not linear and dependent on the exoplanetary context, including stellar class and UV activity \citep{2022A&A...665A.156K}.
    \item Methane (CH$_4$). It is produced by anaerobic metabolism and fermentation of organic matter. Since is highly reactive in an oxidizing atmospheres, it requires a constant production to be detectable. Methane features spectral lines in the UV/VIS/NIR, but these overlap with much stronger water vapor lines.
    \item Nitrous oxide (N$_2$O). This molecule is produced by microbial denitrification of nitrates in oxygen-poor environments \citep{Schwieterman2018}. While nitrous oxide exists only in trace amounts in Earth's atmosphere, it could be more abundant in exoplanets orbiting M-dwarfs. This gas is detectable in the near- and mid-infrared, though its features are weak (at Earth's abundances) and could blend with those of water vapor, carbon dioxide, and methane.
    \item Sulfur-based gases. Hydrogen sulfide (H$_2$S), sulfur dioxide (SO$_2$), carbonyl sulfide (OCS), and dimethyl sulfide (DMS) are potential biosignatures, as they can be metabolic byproducts of sulfur-reducing bacteria and cyanobacteria. Many sulfur compounds have spectral features in the MIR, though they overlap with other gases such as ozone. 
    \item Methylated hydrogens. Methyl chloride (CH$_3$Cl) and methyl bromide (CH$_3$Br) are produced by biological processes, including algae, plants, and decaying organic matter, as well as by some industrial activities on Earth. These compounds have been proposed as capstone biosignatures \citep[see, e.g.,][]{2024RvMG...90..465S,Leung_2022}, metabolic products with high biological activity and low-false positive potential (see Sec.~\ref{subsub:false_positives}). Although not detectable at Earth's abundances, their detection at higher concentration would add supporting evidence for biological activity. These gases are observable in the MIR.
\end{itemize}

Since many, possibly all, of these molecules can also be produced through non-biological processes (see Sec.~\ref{subsub:false_positives}), the interpretation of gaseous biosignatures requires careful consideration. To improve confidence in a detection, scientists look for biosignature pairs that indicate atmospheric disequilibrium conditions, unlikely to persist without continuous biological input. One example is the simultaneous presence of O$_2$ and CH$_4$, which should not coexist for long without continuous biological replenishment \citep{Lederberg, Lovelock}.

When analyzing the reflected spectrum of an exoplanet, we might be able to observe portions of its surface, which will scatter light in specific ways depending on its composition. One of the most well-known surface biosignatures is the Vegetation Red-Edge (VRE), a sharp increase in reflectance around 700 nm caused by the way chlorophyll absorbs and scatters light \citep{2019Galax...7...82C}. The VRE has been observed on Earth and could, in principle, be detected on exoplanets with large landmasses covered in photosynthetic organisms. Future observing facilities working in the visible (HWO, RISTRETTO, PCS) may be able to constrain its presence on nearby rocky exoplanets (e.g. Proxima b). 
However, the challenge with surface biosignatures is that planetary conditions such as cloud cover, oceans, or different types of alien vegetation could alter the expected reflectance signals. 
On that topic, there have been studies aimed at investigating the impact of a different stellar spectra on photosynthetic species \citep[e.g.][]{2020Life...11...10C,2023Life...13.1641B}.

Finally, the temporal biosignatures focus on seasonal or periodic variations in atmospheric composition, which could point to the existance of biological processes. For example, on Earth, the seasonal cycle of photosynthesis and respiration causes detectable fluctuations in CO$_2$ and O$_2$ levels. Similarly, microbial activity can cause methane to vary on seasonal timescales, due to the photolysis of evaporated water in summer, which generates OH that reacts with CH$_4$ \citep{2023ApJ...946...82M}.  Detecting similar trends on exoplanets could provide strong evidence for life, particularly if the variations follow patterns consistent with biological rhythms rather than abiotic processes. However, temporal biosignatures require long-term monitoring, which is currently beyond the capabilities of most exoplanet observatories \citep{2024RvMG...90..465S}.  
Furthermore, 3-D photochemical modeling by \citep{2025PSJ.....6....5B} has shown that, especially for tidally locked exoplanets, seasonal variations in atmospheric composition can also emerge as a consequence of viewing geometry or an eccentric orbit. Hence, interpreting seasonal variations in biosignatures requires a comprehensive understanding of the orbital configuration and its chemical effects.



    
    
    
    
    
    
    
    
    
    
    
    

\subsubsection{False positives}\label{subsub:false_positives}



One of the main challenges when it comes to biosignatures and the detection of {signs of} life beyond Earth is to rule out false-positives. False-positives can be caused by a wide range of processes, from interstellar chemical reactions to geophysical processes, and what makes it challenging is that a biosignature is essentially something that can only be explained by the presence of life, but not by any abiotic processes, yet we do not have a full inventory of abiotic processes in space. 

A famous false positive for life detection on Mars came in the form of small globules of carbonates with isotopic anomalies found in the Martian meteorite ALH84001 (\cite{Treiman03})and enigmatic elongated features resembling terrestrial bacteria but of much smaller size. It is now accepted by most of the community that purely abiotic chemical processes can explain all these observations.

More than 330 molecules\footnote{\url{https://cdms.astro.uni-koeln.de/classic/molecules}} have been detected in space to date \citep{2022ApJS..259...30M}, meaning that these species can all form from abiotic chemical processes in the interstellar medium, and if they were to be considered as potential biosignatures, abiotic contributions need to be properly accounted for to avoid mistaking the presence of the molecule for a sign of life. Two molecules that were previously \citep[e.g.,][]{2016AsBio..16..465S} considered to be ``relatively clean'' biosignatures - i.e. molecular species without a known abiotic production route in the interstellar medium - were methyl chloride (CH$_3$Cl, also called chloromethane) and dimethyl sulfide (CH$_3$SCH$_3$, DMS). Both molecules are abundant on Earth and under terrestrial conditions almost exclusively produced by biological and industrial processes. In 2017, \citet{2017NatAs...1..703F} demonstrated that CH$_3$Cl and tentatively CH$_3$F were present in the gas of the low-mass star-forming region IRAS 16293-2422 using data from the Atacama Large Millimeter Array, and CH$_3$Cl could also be identified in the coma of comet 67P/Churyumov-Gerasimenko from data obtained by the ROSINA instrument suite onboard the Rosetta mission, demonstrating the species' ability to survive accretion into larger bodies in the early solar system. 

It is remarkable that detecting complex molecules like DMS in exoplanet atmospheres is now within the realm of possibility. Recently, \citet{2023ApJ...956L..13M} and \citet{madhusudhan2025new} reported a tentative detection of DMS in the atmosphere of the exoplanet K2-18b, with the latter study claiming a detection significance of 3.4$\sigma$. However, this claim has been called into question by subsequent independent re-analyses \citep{2025arXiv250118477S,taylor2025there}, which suggest that the signal is not robust and may be artefacts of noise, instrument systematics and modelling assumptions - issues that could potentially be resolved with higher spectral resolution. Moreover, \citet{2024ApJ...976...74H} demonstrated that an abiotic source of DMS is also present in the sublimating ices of comet 67P, and more recently, it has also been detected in the interstellar medium towards the galactic center cloud G+0.693–0.027 \citep{2025ApJ...980L..37S}. 
These findings suggest that DMS may no longer be considered to be an irrefutable biosignature.

Molecular oxygen is another molecule that is a strong indicator for life on Earth, however, it can be produced by abiotic processes in space, as evidenced for instance by its presence in comets 67P \citep{2015Natur.526..678B} and 1P/Halley \citep{2015ApJ...815L..11R} as well as towards the $\rho$ Ophiuchi molecular cloud  \citep{2007A&A...466..999L}. 
Oxygen can also accumulate in the atmosphere as a result of atmospheric escape, a process particularly effective for planets around M-dwarfs (see Sec.~\ref{subsub:escape}).

In recent years, a wealth of molecular building blocks of the four macromolecules of life (amino acids, carbohydrates, lipids and nucleic acids) was detected in the interstellar medium or in pristine solar system matter, such as for instance glycolaldehyde \citep{2000ApJ...540L.107H,2012ApJ...757L...4J}, ethanolamine \citep{2021PNAS..11801314R}, or glycine \citep{2016SciA....2E0285A}, demonstrating the presence of abiotic sources of prebiotic species and their precursors, and making the scenario of a clean biomarker a rather unlikely one. Therefore, the detection of a specific molecule alone will hardly suffice as an indicator for the presence of life, but will require transient events or a very careful subtraction of the abiotic background. 

\subsection{How do we look for biosignatures?}\label{sub:how_look}

\subsubsection{Solar System}\label{subsub:how_SS}


In the Solar System, the search for biosignatures can be conducted both in-situ and via remote sensing. This includes the deployment of Martian rovers, sample return missions, and exploration using space probes. PlanetS members have actively contributed to the development of instrumentation and techniques designed to detect potential biosignatures in our Solar System. Here, we focus on the contributions of laser ionization mass spectrometry (LIMS) and spectropolarimetry (SenseLife) to these efforts.


%
%
\paragraph{\textbf{In-Situ Biosignature Detection with Laser Ionization Mass Spectrometery}}

Laser-based mass spectrometry is a powerful technique for detecting various classes of biosignatures, from single microbes to organics. It operates by using a pulsed laser system to desorb or ablate material while simultaneously ionizing it. The resulting ions are then analyzed with a mass analyzer, which determines the mass of the elements and/or molecules present in the sample material removed by the laser pulse. ORIGIN, standing for ORganics Information Gathering INstrument, is a laser ionization mass spectrometer (LIMS) that is operated in desorption mode. This space-prototype instrument features a miniature reflectron-type time-of-flight mass analyzer and employs a pulsed UV laser system for molecular desorption and ionization, enabling the identification of molecular biosignatures \citep{ligterink_origin_2020}. Additionally, LIMS can be operated in ablation mode, allowing, for example, the analysis of single microbes within Martian mudstone \citep{riedo_detection_2020}, isotope ratios \citep{riedo_laser_2021}, and microstructures \citep{tulej_chemical_2015, lukmanov_chemical_2021}.
Investigations conducted by researchers associated with PlanetS have demonstrated ORIGIN’s capability to detect and identify lipids, amino acids, nucleobases, and polycyclic aromatic hydrocarbons \citep{2022PSJ.....3..241B, 2022PSJ.....3...43K, boeren_origin_2025, 2022PSJ.....3...43K}. These organic molecules are deemed of importance for space science and astrobiology as indicators of potential extraterrestrial life, the origins of life, and prebiotic chemistry \citep{national_academies_of_sciences_engineering_and_medicine_origins_2022}. ORIGIN’s ability to detect a broad range of organic compounds combined with its compact design, highlights its strong suitability for biosignature detection in a variety of environments, exemplified by its potential role in detecting organics in Venus’s cloud layers as part of the Morning Star Mission \citep{2022Aeros...9..312L}.
For a more detailed discussion on laser-based mass spectrometry for space exploration, including specifics on the ORIGIN setup, refer to Chapter XXX by Riedo et al.

\paragraph{\textbf{Our Earth and Moons with SenseLife}}

The SenseLife Project is a follow-up of the former PlanetS project MERMOZ and deals with the remote detection of narrow-banded circular polarization features as biosignatures. Induced circular polarization through scattering of originally unpolarized light (like disc-integrated star light) is a proxy of asymmetric macromolecules which in turn can only be assembled by life from a homochiral environment.

Homochirality is considered to be a unique and universal characteristic of life. Life requires functional macromolecules like DNA/RNA or proteins to store information and to perform metabolic processes. These macromolecules are synthesized from smaller building blocks, for example amino acids for proteins and nucleotides for DNA/RNA. These building blocks often are chiral, they have a left- and a right-handed version called \textsc{l}- and \textsc{d}-enantiomers. For macromolecular assemblages, they are combined to polymers that curl and fold according to the structures and charges of the single building blocks, and thus form the functional macromolecule. If only one part of the polymer sequence would be the opposite enantiomer, the whole molecule could be rendered dysfunctional. Therefore, life for example only implements \textsc{l}-amino acids in proteins and \textsc{d}-sugars in the backbone of the DNA. This enantiomeric composition is also reflected in the environment inhabited by life, but whether life arose from an environment with an already existing enantiomeric excess or whether life itself created homochirality from a completely chaotic mix is still debated.

\begin{figure}
    \centering
    \includegraphics[width=0.7\linewidth]{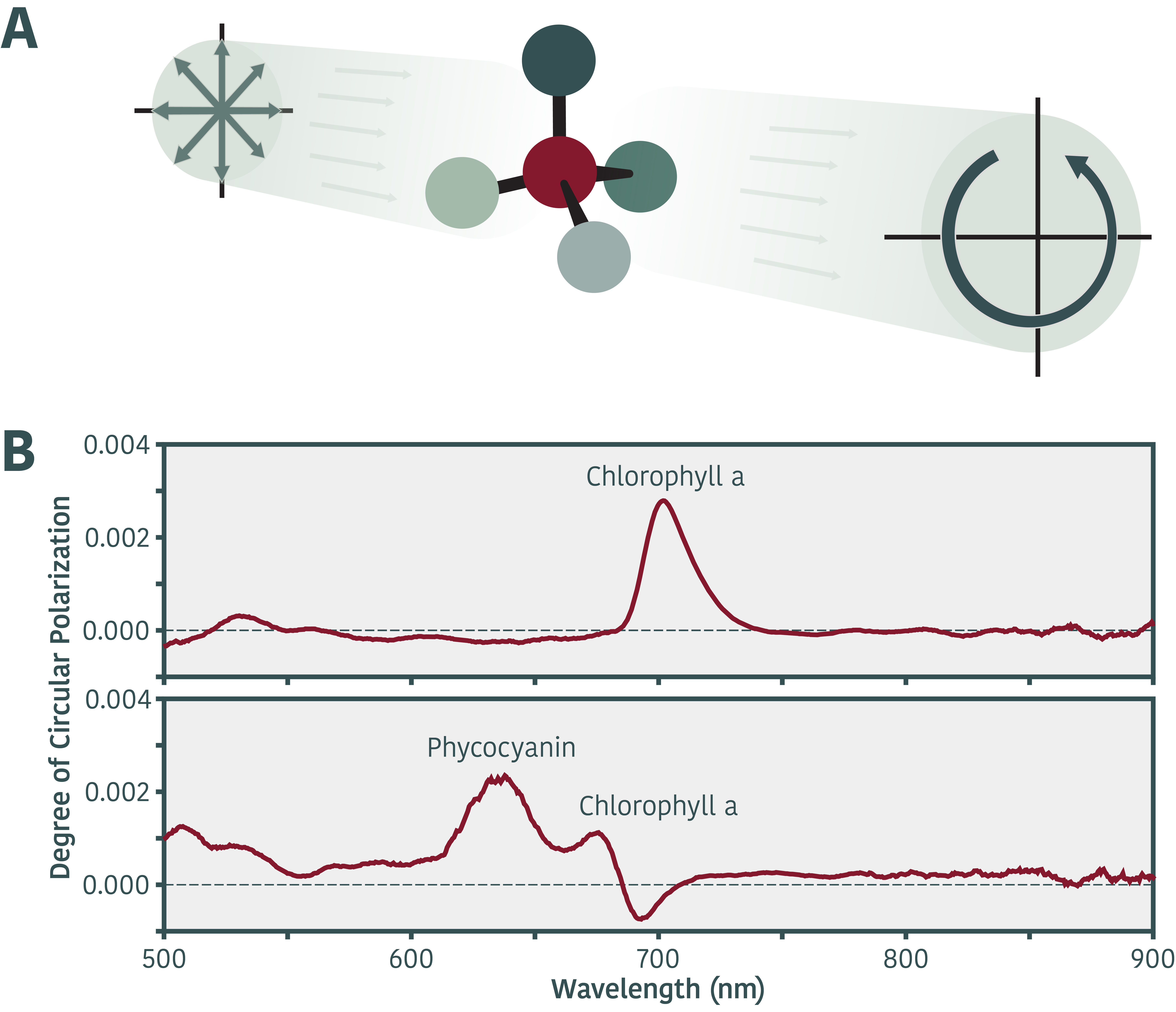}
    \caption{\textbf{A} Illustration of the principle of induced circular polarization. Chiral macromolecules can change the polarization of scattered light to a fraction of circular polarization. \textbf{B} Examples of circular polarization spectra. Upper panel: a leaf in reflectance. Lower panel: a \textit{Synechococcus} (cyanobacteria) culture in transmittance. Pigment types can be recognized as circular polarization peaks in the respective absorption bands. Figure from \citet{2025arXiv250703819B}.}
    \label{fig:SenseLife}
\end{figure}

Either way, homochirality is needed for organic life as we know it, and thus would be a suitable characteristic to search for in the universe. Apart from well known measurement techniques \textit{in situ}, macromolecules constructed from homochiral building blocks can be detected remotely. These asymmetric molecular assemblages preferentially absorb either left- or right-turning circularly polarized light, causing a slight bias in originally randomly mixed (i.e. unpolarized) light (see Fig. \ref{fig:SenseLife}A). This induced partial circular polarization is narrow-banded, the wavelength is depending on the absorbance bands of the respective molecules (see Fig. \ref{fig:SenseLife}B).

For this purpose, SenseLife developed the full-Stokes spectropolarimeter FlyPol which was designed for both, laboratory measurements and air-borne operation in the field between 450 and 850 nm (see Chapter ``Active Moons of the Solar System'' for technical details). FlyPol showed to be capable to differentiate between different surface environments with vegetation (forests, meadows, lakes) and abiotic surfaces (roofs, streets) purely based on their polarimetric signature, both from a helicopter and a hot-air balloon. In the near future, FlyPol will be performing higher flights on an aircraft and be implemented in the 1-m telescope SAINT-EX (cross-reference). On the mid term, FlyPol may perform experiments from the International Space Station and be implemented in a cube sat orbiter. On the long term, we plan to employ a spectropolarimeter on missions to the icy moons to identify areas of astrobiological interest on the surface to be later explored \textit{in situ} by a lander.

\subsubsection{Exoplanets}\label{subsub:how_exo}

%
%
%
%
%
%

PlanetS has been heavily involved in the development of many instruments aiming at detecting biosignatures (RISTRETTO, ANDES, PCS from the ground and LIFE from space) and detecting targets for these instruments (ESPRESSO, NIRPS, SPECULOOS... but we here focus on SAINT-EX). 
These instruments will likely shape the future of temperate exoplanets science.

\paragraph{\textbf{The SAINT-EX Observatory}}
The SAINT-EX Observatory is a fully robotic facility that hosts a 1 meter Ritchey-Chretien telescope. The main mission of Saint-Ex is the search for terrestrial exoplanets orbiting ultracool stars similar to the Trappist-1 system.

Finding exoplanets around ultracool dwarf stars is a baby step towards the detection of life as we know it out of Earth and out of the Solar system. The search for {signs of }life out of the solar system should begin with knowing where to begin such a search.
For a guarantee that a planet will evolve along a path similar to our Earth, it must be given enough time to follow this slow process of evolution and life formation. While Sun-like stars take about ten billion years to transform into red giants and consume any planet existing close to them, the time span for this evolution in ultra-cool dwarfs is over one hundred billion years.\\
Another important reason why the search for life around ultra-cool dwarf stars has the tendency to produce results is the fact that they are extremely abundant in our galaxy and hence are more within reach than other stellar counterparts. It is a known fact that seventy percent of stars in the Milky Way are of the M-dwarf type \citep{henry2024character}.
Having this ability to detect exoplanets orbiting ultracool stars would lead to the detection of a good number of planets in the habitable zone of their host stars. These are therefore planets with the capability to host life as we know it. Data from SAINT-EX contributed to the recent validation of the Earth-sized explanet (SPECULOOS-3b) orbiting an ultra-cool red dwarf very close to us (55 light years) \citep{gillon2024detection}.

The Saint-Ex observatory will host a polarimeter instrument in the near future, this instrument will be able to detect circularly polarized light from observed bodies, and this would be a step toward detecting homochirality from extraterrestrial bodies.

In addition to the observation of M-type stars hosting transiting planets, there is a project to observe a number of brown dwarfs that could be serving as hosts to transiting planets. Brown dwarves, because of their low mass and low temperature nature, could serve as a sort of low-hanging fruit for the search for {signs of} life out of the solar system \citep{Bolmont2025}. These objects have the habitable zone quite close to themselves and, according to Kepler's laws, orbiting planet-sized objects should be quite close to the host. This would lead to low periods of revolution and hence Saint-Ex would seize such an opportunity to collect large amounts of transit data within a short time.

\paragraph{\textbf{High resolution spectroscopy from the ground: RISTRETTO, ANDES, PCS}}



Thanks to transmission and emission spectroscopy, the James Webb Space Telescope (JWST) has ushered in a new era for atmospheric characterization of temperate terrestrial exoplanets, such as the Trappist-1 system planets \citep{2023Natur.618...39G,2023Natur.620..746Z,2025arXiv250902128G}. 
However, these methods remain restricted to transiting exoplanets, which are not the most frequent due to the relatively rare transit geometry. 
A new method combining high-contrast imaging and high-resolution spectroscopy in reflected light (\citealt{snellen2015, lovis2017} and chapter instrumentation) will soon allow upcoming ground-based instruments---such as RISTRETTO \citep{2024SPIE13096E..1IL}, ANDES \citep{2024SPIE13096E..13M,2025ExA....59...29P} and PCS \citep{2013aoel.confE...8K,2021Msngr.182...38K}---to enable atmospheric characterization of nearby non-transiting terrestrial exoplanets that were previously not accessible. This will offer a larger sample of terrestrial exoplanets to explore, providing a powerful alternative to JWST.

RISTRETTO (high-Resolution Integral-field Spectrograph for the Tomography of Resolved Exoplanets Through Timely Observations) will be proposed as a visitor instrument at the Very Large Telescope (VLT). The primary science goal of RISTRETTO is to detect and characterize atmospheres of nearby exoplanets in reflected light for the first time. In particular, it will focus on the detection of the reflected light from Proxima Centauri b in the visible, the closest exoplanet from the Earth \citep{2016Natur.536..437A}, at a contrast of about 10$^{-7}$. The required exposure time is estimated at $\sim$50 hours for a 5$\sigma$ detection, allowing to infer the orbital inclination, the true mass and the albedo \citep{bugatti_ristretto}.

RISTRETTO will pave the way for the development of similar high-resolution spectrographs at the Extremely Large Telescope (ELT), such as the second-phase instrument ANDES 
and the potential third-phase instrument PCS, 
that will broaden the spectral range explored by covering all the visible and near infrared. The overarching vision driving the instrumental design of ANDES and PCS is to continuously enhance their performance by achieving unprecedented angular and spectral resolutions, detecting reflected light atmospheric signals in shorter timeframe and overcoming current limitations in atmospheric characterization of Earth-like exoplanets. 

For ANDES, a ``golden sample'' of 5 targets around nearby M dwarf stars was identified for atmospheric characterization: Proxima~b, GJ 273~b, Wolf 1061~c, GJ 682~b and Ross 128~b. The exposure times required to detect each planet's atmosphere with ANDES are respectively 0.67, 6.5, 5.8, 7.2 and 13 nights (assuming an albedo of $\sim$0.3, \citealt{2025ExA....59...29P}). Specifically, Proxima~b can be characterized in a single night of observations ($\sim$8 hours) at quadrature (corresponding to the maximum angular separation) and in significantly less time when it is closer to superior conjunction.


These initial estimates were based on an ad hoc planetary albedo, but refining predictions for upcoming observations requires a more accurate assessment of the planets' spectral reflectivity. A promising approach is to conduct 3-D climate simulations across various configurations, analyzing potential observable signatures for each scenario. This will help evaluate their detectability and determine the capability of instruments to constrain their actual climates.

During the past few years, \cite{Houelle_Ross128b}, \cite{Houelle_Proximab} has focused on two cases: Ross 128~b \citep{2018A&A...613A..25B} and Proxima~b \citep{2016Natur.536..437A} and has performed climate simulations using the Generic PCM (see Sec.~\ref{sub:Tools_habitability}) to 1) investigate the planets' habitability and 
2) to compute synthetic observables using the radiative transfer codes Pytmosph3R \citep{2022A&A...658A..41F} and PICASO \citep{2019ApJ...878...70B}. 
These codes can compute synthetic reflected light spectra and phase curves based on GCM simulations. 
This way they capture 
information about: the distribution and properties of clouds, the molecular composition and properties of the atmosphere (scattering properties), the surface composition and potential vegetation, and the potential presence of an ocean (glint). 3-D modeling is vital to estimate accurately the observability of such faint targets.

Ross 128~b  is located 3.4 parsecs from Earth, is the second closest known Earth-like planet and the closest relatively temperate exoplanet orbiting a quiet star. For most of the cases explored in \cite{Houelle_Ross128b} (different atmospheric compositions, surface pressures, greenhouse gas concentration, global water inventories and rotational configurations), Ross 128~b might be too irradiated to permanently sustain liquid water on its surface, especially on its dayside. This would cause the planet to have a low albedo, which is not favorable for its potential habitability and detectability.  

\begin{figure}
    \centering
    \includegraphics[width=0.9\linewidth]{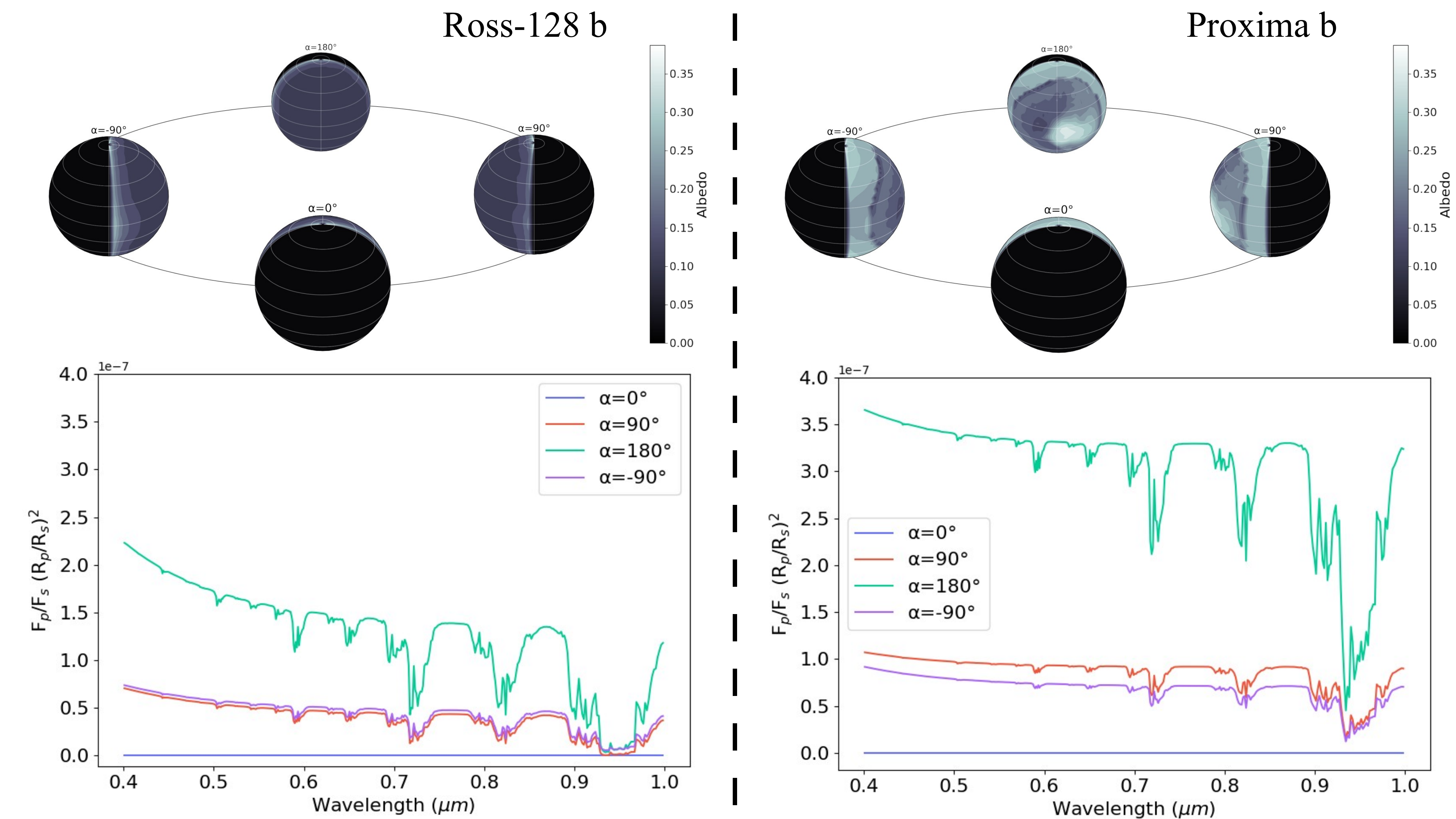}
    \caption{Albedo maps and reflected spectra of Ross 128~b (left) and Proxima~b (right).Top row: albedo maps for 4 geometrical configurations. Due to the presence of ice near the terminator and clouds on the dayside, Proxima~b has a higher albedo than Ross 128~b. Bottom row: corresponding reflected light spectra computed with Pytmosph3R.}
    \label{fig:ReflectedLight}
\end{figure}

However, Proxima~b, located at $\sim$4.2 light-years from the Earth, and the closest potentially habitable exoplanet from us, has been shown to be potentially habitable \citep{2016A&A...596A.112T,2019AsBio..19...99D}. 
The goal of \cite{Houelle_Proximab} is to revisit the results of \citep{2016A&A...596A.112T,2019AsBio..19...99D} with a new, computationally efficient ocean model which is a compromise between the simplified ocean model employed in \citet{2016A&A...596A.112T} and a full 3-D ocean model as in \citet{2019AsBio..19...99D} (\citealt{bhatnagar_ocean}, see Sec.~\ref{sub:Earth_habitability}). 
Preliminary results shown in Figure~\ref{fig:ReflectedLight} highlight that Proxima~b is a much better target than Ross 128~b in terms of habitability and observability.


\paragraph{\textbf{The Large Interferometer For Exoplanets (LIFE)}} \label{paragraph:LIFE}
The LIFE mission is a proposed space observatory designed to detect and analyze the atmospheres of exoplanets\footnote{\url{www.life-space-mission.com}}. As a mid-infrared nulling interferometer, LIFE will combine light from multiple apertures (separated by tens to hundreds of meters)  to achieve the required  spatial resolution and  measure the intrinsic thermal emission of the exoplanets. It is optimized for detecting life beyond the Solar system and its unique capabilities exceed those of all existing and currently planned space missions and ground-based observatories. LIFE will focus on stars in the immediate Solar neighborhood and directly detect hundreds of exoplanets, 30 to 50 of which will be similar in size and temperature to Earth. LIFE will make a detailed inventory of their atmospheric composition and constrain their sizes, temperatures, and atmospheric conditions. As such, LIFE can identify biospheres on exoplanets that are like Earth's via well-known biosignature gases, but LIFE can also find atmospheric biosignatures from biospheres that differ significantly from Earth's in their composition or stellar environment. Even more, LIFE can also find imprints of extraterrestrial technology, such as artificial green-house gases, in the atmospheres of exoplanets. 

LIFE has been conceived and led by colleagues from PlanetS in Switzerland and  has acquired a large international support base. The science capabilities of LIFE have been presented and discussed in a number of papers. These include
\begin{itemize}
    \item A general introduction to the LIFE mission, the LIFE simulator, and detection yield estimates \citep{2022ExA....54.1197Q,2022A&A...664A..21Q,2022A&A...664A..22D,2024SPIE13095E..1FH,2022A&A...668A..52K}.
    \item Atmospheric retrieval studies of Earth- and Venus-like exoplanets \citep{2022A&A...665A.106A,2022A&A...664A..23K,2023A&A...673A..94K}.
    \item Investigations about the detectability of non-traditional atmospheric biosignatures and technosignatures \citep{2023AsBio..23..183A,2024AJ....167..128A,2025ApJ...982L...2L, 2024ApJ...969...20S}.
    \item The demonstration that LIFE can correctly quantify Earth-like biosignatures in data obtained from Earth observing satellites and identify Earth as a habitable and inhabited planet \citep{2023ApJ...946...82M,2024ApJ...963...24M,2024ApJ...975...13K}.
    \item A quantitative assessment showing that LIFE can detect many more of the already known nearby exoplanets (i.e., detected by RV surveys) than a space mission aiming at detecting these planets in reflected light \citep{2023A&A...678A..96C}.   
\end{itemize}
In addition, the technical aspects of LIFE and related R\&D efforts have been gaining significant traction. A first updated baseline architecture was recently presented in \cite{2024SPIE13095E..1DG}, but earlier papers already looked at different architecture designs \citep{2022A&A...664A..52H}.
Also, ideas for pathfinder and tech-demonstration missions are being discussed \citep[e.g.,][]{2024SPIE13095E..1GH}.
Finally, significant progress was also made in demonstrating the LIFE measuring principle in the laboratory \citep{2024SPIE13095E..1HR,2024SPIE13095E..38B} and investigations of the potential use of mid-infrared photonic devices in LIFE is ongoing \citep{2024SPIE13100E..6OM}.

LIFE is still in the concept and feasibility study phase, but scientists and engineers are working on refining the mission design and developing the necessary technology with increasing momentum. Ultimately, the unprecedented capabilities of the LIFE mission have the potential to fundamentally transform our understanding of our place in the Universe by systematically searching for signatures of life beyond the Solar System.

\section{Summary}\label{sec:summary}

Understanding the climates of terrestrial exoplanets and the detectability of biosignatures is an inherently interdisciplinary challenge, requiring the integration of insights from Solar System exploration, exoplanet observations, and climate science. Building from the only known inhabited planet, Earth, NCCR PlanetS researchers have developed models, tools, and observational strategies to assess planetary environments far beyond direct reach.

Between 2018 and 2025 (phases 2 and 3), PlanetS has made major contributions to this field. On the modelling side, the development and application of the Generic Planetary Climate Model (Generic-PCM) enabled robust climate studies across a wide range of planetary regimes, from early Venus to temperate terrestrial exoplanets, incorporating advanced modules such as a dynamical slab ocean. These and related efforts provide key constraints on planetary habitability and climate stability. In parallel, THOR, another global climate model designed specifically for exoplanetary studies, was developed to avoid Earth-centric assumptions and to stably simulate diverse atmospheric regimes.

PlanetS researchers have also advanced atmospheric retrieval techniques, combining forward-modelling, Bayesian inference, and machine learning to interpret observations ranging from pictures of the surface of Europa, to phase curves and directly imaged spectra of exoplanets. These efforts have been central to assessing scientific return and refining instrumental requirements for future missions, notably LIFE, demonstrating its ability to detect Earth-like atmospheres and biosignatures.

Within the Solar System, PlanetS contributed instrumentation and methodologies for biosignature detection, including laser ionization mass spectrometry (ORIGIN) and spectropolarimetry (SenseLife), enabling in-situ and remote detection of organics, isotopic ratios, and microstructures relevant to astrobiology. These tools extend our ability to probe planetary surfaces, ices, and atmospheres for signs of prebiotic chemistry or life.

Finally, PlanetS has been instrumental in preparing the next generation of observatories, from JWST, VLT and ELT instruments such as RISTRETTO and ANDES, to LIFE and the Habitable Worlds Observatory. Through simulations, retrieval frameworks, and mission studies, PlanetS has helped define observational strategies that maximize our chances of detecting habitable, and possibly inhabited, worlds.

In summary, PlanetS contributions span theory, instrumentation, and mission design, creating an integrated framework to advance the search for life beyond Earth. By bridging Solar System and exoplanet science, PlanetS has significantly advanced our ability to model climates, assess habitability, and recognise biosignatures. This ensures that when the first robust evidence of extraterrestrial life is found, PlanetS will have played a central role in making it possible.




\begin{acknowledgement}
This work has been carried out within the framework of the NCCR PlanetS supported by the Swiss National Science Foundation under grants 51NF40\_182901 and 51NF40\_205606. 
The climate computations (for Fig.~\ref{fig:ReflectedLight}) were performed at the University of Geneva's HPC clusters of Baobab and Yggdrasil. This review has made use of the Astrophysics Data System, funded by NASA under Cooperative Agreement 80NSSC25M7105. 
EB acknowledges the financial support of the SNSF (grant number: 200021\_197176 and 200020\_215760). 
SFW acknowledges the financial support of the SNSF Eccellenza Professorial Fellowship (PCEFP2\_181150) and helpful discussions with the University of Bern's ROSINA team on the abiotic molecular inventory of cometary ices.
GC acknowledges the financial support of the SNSF (grant number: P500PT\_217840).
KH acknowledges funding from the Belgian Federal Science Policy Office's FED-tWIN research program STELLA (Prf-2021-022).
EA’s work has been partly carried out within the framework of the NCCR PlanetS supported by the Swiss National Science Foundation under grants 51NF40\_182901 and 51NF40\_205606. 
EA’s research was partly supported by an appointment to the NASA Postdoctoral Program at the NASA Goddard Space Research, administered by Oak Ridge Associated Universities under contract with NASA (ORAU-80HQTR21CA005).
JH thanks the Center for Origin and Prevalence of Life (COPL) at ETH Zurich for their support.
KH acknowledges the support of European Research Council ERC-2017-CoG-771620 EXOKLEIN.
\end{acknowledgement}

\tiny
\bibliographystyle{spbasic.bst}
\bibliography{bibliography}{}

%

%
%
%
%

\end{document}